\documentclass{article}

\usepackage{microtype}
\usepackage{graphicx}
\usepackage{subfigure}
\usepackage{booktabs} % for professional tables

\usepackage{xcolor}

\usepackage{hyperref}

\usepackage[accepted]{icml2025}
\usepackage{amsmath}
\usepackage{amssymb}
\usepackage{mathtools}
\usepackage{amsthm}

\usepackage[capitalize,noabbrev]{cleveref}

\theoremstyle{plain}

\theoremstyle{definition}

\theoremstyle{remark}

\usepackage[textsize=tiny]{todonotes}

\begin{document}

\twocolumn[
%\icmltitle{Multi-Modal Foundational Model for AI-Driven Physical-Layer Wireless Communication: Toward Intelligent NextG Networking}

\icmltitle{A Multi-Modal Foundational Model for Wireless Communication and Sensing}




\begin{icmlauthorlist}
\icmlauthor{Vahid Yazdnian}{}
\icmlauthor{Yasaman Ghasempour}{}
\end{icmlauthorlist}

\begin{center}
Department of Electrical and Computer Engineering, Princeton University, USA
\end{center}

%\icmlaffiliation{yyy}{Department of Electrical Engineering, Princeton University, USA}
%\icmlaffiliation{comp}{Company Name, Location, Country}
%\icmlaffiliation{sch}{School of ZZZ, Institute of WWW, Location, Country}

%\icmlcorrespondingauthor{Firstname1 Lastname1}{first1.last1@xxx.edu}
%\icmlcorrespondingauthor{Firstname2 Lastname2}{first2.last2@www.uk}

\icmlkeywords{AI-Driven Physical-Layer Wireless Communication, Foundational Model}

\vskip 0.3in
]





\begin{abstract}
\vspace{-1mm}
Artificial intelligence is a key enabler for next-generation wireless communication and sensing.  Yet, today’s learning-based wireless techniques do not generalize well: most models are task-specific, environment-dependent, and limited to narrow sensing modalities, requiring costly retraining when deployed in new scenarios. This work introduces a task-agnostic, multi-modal foundational model for physical-layer wireless systems that learns transferable, physics-aware representations across heterogeneous modalities, enabling robust generalization across tasks and environments.
Our framework employs a physics-guided self-supervised pretraining strategy incorporating a dedicated physical token to capture cross-modal physical correspondences governed by electromagnetic propagation. The learned representations enable efficient adaptation to diverse downstream tasks, including massive multi-antenna optimization, wireless channel estimation, and device localization, using limited labeled data. Our extensive evaluations demonstrate superior generalization, robustness to deployment shifts, and reduced data requirements compared to task-specific baselines. %We further provide a large-scale multimodal dataset and standardized benchmarks to support systematic development and evaluation of physical-layer wireless foundational models. 

\end{abstract}

\vspace{-9mm}
\section{Introduction}
\label{intro}
\vspace{-2mm}
%%maybe this
%Over the past decade, artificial intelligence (AI) has transformed numerous fields, including wireless communications, where it has evolved from a supportive tool into a central driver of innovation. This evolution leaves little doubt that AI will play a pivotal role in future mobile wireless networks.
To meet the ever-increasing demand for higher data rates, wireless systems are becoming dramatically more complex: adoption of large-scale antenna arrays, operation at millimeter-wave and sub-terahertz frequencies, and tight integration of communication and sensing are fundamentally reshaping the physical layer.
As spectrum utilization becomes denser and applications (e.g, AR/VR) increasingly delay-sensitive, AI-driven optimization and adaptation become increasingly important and have already demonstrated promising results in complex physical-layer tasks, ranging from beam management~\cite{brilhante2023literature} and channel state information (CSI) prediction ~\cite{liu2025wifo,fan2025csi} to multiple-input multiple-output (MIMO) precoding~\cite{kamal2025hybrid}, and wireless positioning within communication networks~\cite{demirhan2023integrated,wang2024generative}. These advances have established AI-driven physical-layer wireless communication as a central driver of innovation and a cornerstone of next-generation networking~\cite{ye2024artificial,chataut20246g,lin2025bridge}. 
%%maybe this
%This trend is reflected in both academia and industry, with the 3rd Generation Partnership Project actively investigating AI-based CSI feedback, beam management, and positioning for potential inclusion in future 6G standards~\cite{akyildiz20206g,lin2025bridge}.

Despite these advances, most conventional solutions rely on domain-specific solutions, which are often task- and site-specific, limiting generalization across scenarios and necessitating expensive on-site data collection and re-training for separate downstream models. In contrast, the practical integration of AI into the core of wireless systems calls for a fundamentally new paradigm, one capable of addressing the key limitations of classical deep learning approaches: scalable acquisition of representative data, robust generalization across deployment sites, reliable inference under stringent latency constraints, and the incorporation of transparent, physics-grounded inductive biases into the learning process. 

These challenges fundamentally stem from the fragmented nature of existing learning-based wireless solutions, which are typically trained on narrow data modalities~\cite{yang2025wirelessgpt} and/or optimized for isolated tasks~\cite{lee2024llm,guo2025prompt}. \textit{Wireless systems, however, are inherently multimodal physical systems, where radio-frequency measurements, spatial geometry, and environmental structure are tightly coupled through the laws of electromagnetic (EM) wave propagation.} The path forward, therefore, demands models that not only learn \textit{from} network data, but also learn \textit{with} the underlying physics that governs it.
%Multimodal foundational models provide a natural framework to meet this need by jointly learning the universal representations from heterogeneous data sources. 
%Such representations can enable strong generalization across tasks and environments, while supporting efficient adaptation to new downstream applications with limited additional supervision, eliminating the need for extensive task-specific on-site data collection.

\begin{figure}[t]
    \centering
\includegraphics[width=0.48\textwidth]{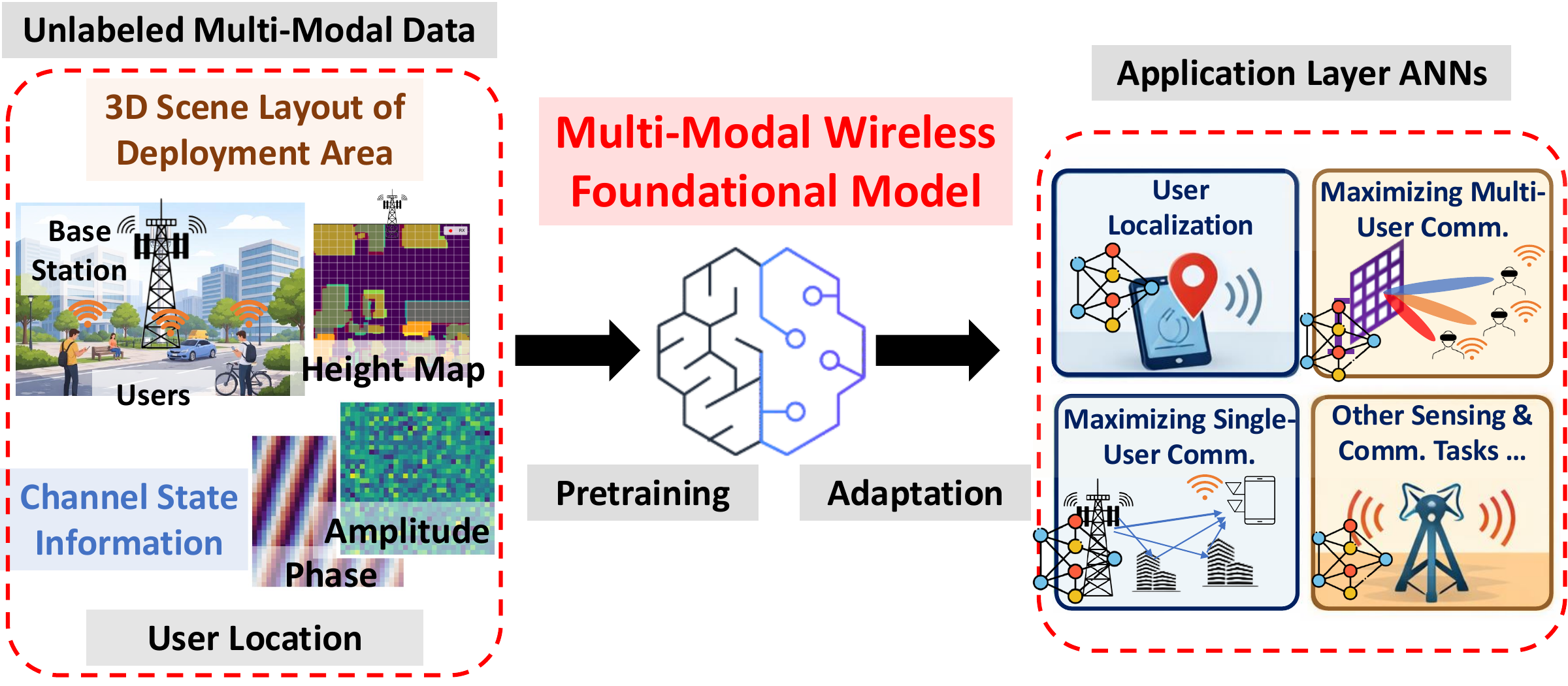} 
\vspace{-8mm} 
\caption{\small Foundational models serving application layer Artificial Neural Networks (ANNs) in wireless communication and sensing.}
\vspace{-6mm} 
    \label{fig:overview}
\end{figure}

To meet this need, we present a multi-modal foundational model for AI-driven physical-layer wireless systems, offering a principled pathway to capture the shared physical correspondence inherent in such systems through physics-guided pretraining, grounded in EM propagation principles, on large-scale, heterogeneous data spanning multiple sensing and communication modalities, as illustrated in Figure~\ref{fig:overview}. By learning transferable physics-aware representations, our framework enables efficient adaptation to many wireless downstream tasks that are fundamentally difficult to learn, delivering strong performance with limited additional supervision and thereby eliminating the need for extensive task-specific on-site data collection.

At the core of our proposed unified framework, we propose a physics-guided feature embedding framework leveraging transformer architectures to capture global, robust representations of the wireless channel. The model learns cross-modality associations among the three-dimensional environment surrounding the communicating parties, the spatial locations of the transmitter and receiver, and the temporal–spectral characteristics of the wireless signals observed at the receiver.
We devise a physics-informed self-supervised pre-training procedure, tailored to the physical correspondence among heterogeneous input modalities, consisting of two key components that enable robust and transferable feature embeddings: \textit{(i)} \textbf{Reconstruction Loss Minimization}, where the model learns the physical associations among different modalities by generating embeddings from multiple viewpoints that enables the reconstruction of masked portions of the input data; and \textit{(ii)} \textbf{Physics-Guided Cross-Modal Distillation}, where we introduce a dedicated $[\mathrm{PHYSC}]$ token as a global aggregator to inject the underlying physically meaningful cross-modal correspondence, represented as \emph{spatial spectra}, which describe the distribution of signal energy across spatial directions. %This is achieved by guiding the model to embed relevant physical information so as to minimize the reconstruction error of the targeted physical property.

While Large Language Models (LLMs) benefit from mature benchmarks and datasets that accelerate architectural development and evaluation, no comparable framework exists in wireless systems. This work bridges that gap by introducing not only the foundational model, but also the supporting tools, pretraining dataset, and benchmarking infrastructure needed for systematic implementation, evaluation, and deployment, enabling continued progress and broader adoption by the research community. Within this unified framework, we evaluate the introduced wireless foundational model across a range of key benchmarking tasks envisioned as critical enablers for next-generation wireless systems~\cite{akyildiz20206g}, including data-rate optimization through hybrid multi-input-multi-output (MIMO) precoding in both single-user and multi-user communications, as well as wireless localization from partial channel state information. In particular, we compare performance (e.g., achievable sum data rate, localization error), and domain-specific data requirements against task-specific models. Our results demonstrate the proposed foundational model’s superior transferability to new applications and scene properties, and its ability to achieve strong key performance indicators (KPIs) with minimal on-site, task-specific data collection.

%Our contributions include: 1) we present the first task-agnostic multimodal foundational model for physical-layer wireless systems, which exhibits strong generalizability by transferring learned representations to diverse downstream applications and new deployment sites, 2) we propose a novel physics-guided self-supervised pretraining framework specifically tailored for multimodal wireless foundational models, 3) we develop a large-scale multimodal physical-layer wireless dataset covering 10,000 realistic outdoor urban environments, enabling the scalable training of foundational models, 4) we formulate key benchmarking protocols and standardized evaluation datasets to support systematic performance evaluation and promote broader adoption within the research community.

%\vspace{-3mm}
\section{Related Work}
\label{related_work}
%\vspace{-2mm}
%elbir2019hybrid,letaief2019roadmap
There is rich literature on applying conventional data-driven learning-based methods for optimizing wireless network performance, targeting individual tasks under fixed assumptions and deployment conditions~\cite{huang2020reconfigurable,ozpoyraz2022deep}. Unfortunately, such architectures struggle to generalize and incur significant costs in data acquisition and model re-training, making them difficult to deploy in practice. Motivated by these limitations, recent work has been increasingly exploring foundational models for wireless systems~\cite{jiang2025towards, yu2025channelgpt, guo2025large, mengistu2026learning}. Existing studies in this emerging area can be categorized into two main directions.

%\vspace{-1mm}
\noindent\textbf{(1) Task-Specific Frameworks:} 
A line of recent works focuses on adapting pretrained large-scale models, primarily LLMs, to solve specific wireless downstream tasks or a small set of closely related tasks. In wireless communications, such models have been applied to resource allocation~\cite{lee2024llm}, CSI prediction and feedback~\cite{liu2024llm4cp,guo2025prompt}, and spectrum management~\cite{zhou2025spectrumfm}, demonstrating improved generalization and reduced data requirements in new scenarios. In wireless sensing, works such as~\cite{li2025large, pan2025large} integrate LLMs for localization and integrated sensing and communication in UAV-assisted networks, achieving enhanced robustness under limited labels and adaptability to unseen base station configurations. In a related direction,~\cite{liu2025llm4wm,zheng2025large} propose the use of a mixture-of-experts based on LoRA-adapted LLMs as a shared backbone to handle predefined wireless tasks, while~\cite{zheng2025muse} leverages LLMs to guide the encoding of task-dependent input information within a unified transformer-based architecture, enabling support for a set of closely related wireless communication tasks. Despite promising performance, the aforementioned approaches face several fundamental limitations. First, they are largely task-specific, limiting generalizability and hindering the learning of universal representations transferable across diverse physical-layer tasks. In addition, most existing methods rely on pretrained LLMs not originally trained on physical-layer wireless modalities, causing domain and knowledge shifts that can degrade performance when adapting to wireless-specific data distributions.

%These limitations highlight the need to move toward the development of explicitly tailored, generalizable foundational models for physical-layer wireless systems that learn universal representations grounded in the underlying physics.

\noindent\textbf{(2) Task-Agnostic Frameworks:} 
In contrast, recent works~\cite{alikhani2024large, yang2025wirelessgpt} have presented early efforts toward task-agnostic wireless channel foundational models. These approaches focus on single-modality channel representations and employ self-supervised pretraining to learn expressive CSI feature extractors, demonstrating performance gains over conventional model-based techniques and smaller learning-based models, particularly in regimes with limited labeled data. However, existing task-agnostic approaches exhibit limitations. They focus exclusively on a single modality, which can induce distribution shifts at inference time when deployed in unseen environments, thereby hindering robust adaptation. Further, owing to their pretraining paradigms, these models are typically optimized for narrowly defined single-modality reconstruction objectives, limiting their ability to capture the shared underlying physical structure that governs sensing and communication problems. As a result, their representations remain insufficient to support unified physical-layer intelligence across diverse wireless tasks and scenarios.

%In this work, we introduce the first task-agnostic \emph{multi-modal foundational model} for physical-layer wireless systems, leveraging physics-guided self-supervised pretraining and modality selection. By jointly learning from heterogeneous modalities, our model acquires universal representations that transfer effectively across diverse sensing and communication tasks and unseen deployments, advancing toward unified physical-layer intelligence in next-generation networks.

\begin{figure*}[t]
    \centering
\vspace{-2mm} 
\includegraphics[width=0.85\textwidth]{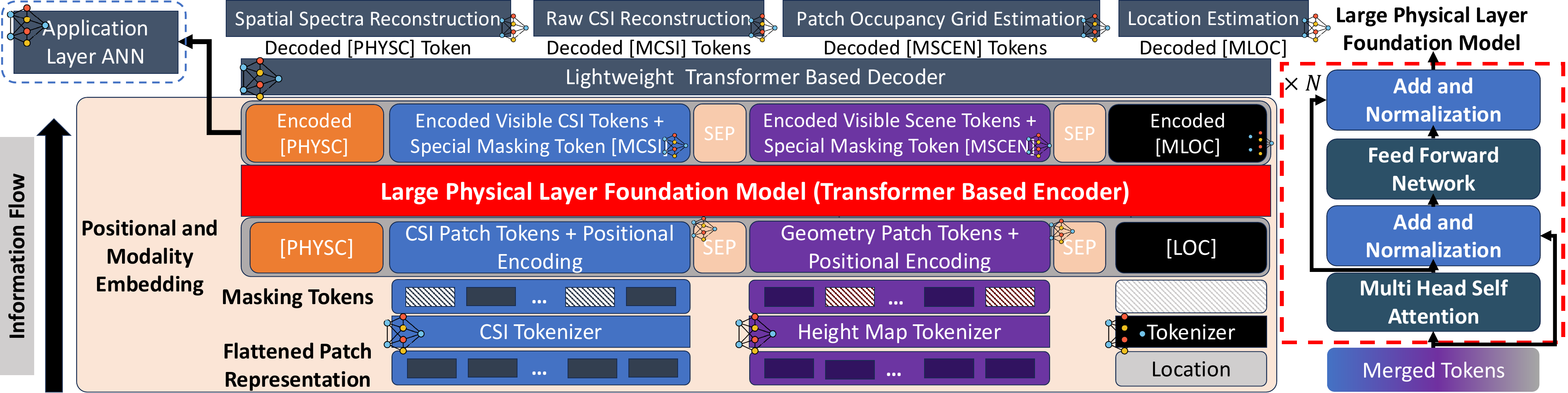} 
\vspace{-4mm} 
\caption{\small Overview of the proposed multi-modal foundation model architecture and its physics-guided end-to-end learning pipeline.}
\vspace{-5mm}
    \label{fig:main_abstract_img}
\end{figure*}
%\vspace{-3mm}

\section{Multi-Modal Foundational Model}
%\vspace{-1mm}
\subsection{Preliminaries and Fundamental Design Decisions}
\vspace{-1mm}
Physical-layer wireless systems inherently involve modalities and structural dependencies that are fundamentally different from those encountered in conventional pre-trained LLMs or VLMs, as they are governed by electromagnetic propagation characteristics. Consequently, a fundamentally different modeling paradigm is required: first, to identify the relevant interacting modality types, and second, to design training strategies that explicitly adhere to the physical properties of the underlying problem. 

Hence, in designing foundational models tailored to physical-layer wireless systems, it is essential to explicitly adhere to the fundamental properties of the wireless channel between the transmitter (base station) and receiver (user equipment), particularly in MIMO settings with multiple transmit ($N_t$) and receive ($N_r$) antennas and corresponding radio-frequency chains. An MIMO wireless channel can be theoretically characterized by the fundamental laws of EM wave propagation and reflection in the following form:
\textcolor{black}{
\begin{equation}
\label{eq:wireless_channel}
\scalebox{0.9}{$
H_{i,j}[f] = \sum_{\ell=1}^{L} \alpha_\ell \, 
a_{r,i}(\theta_\ell^{(r)}, \phi_\ell^{(r)}) \, 
a_{t,j}^{*}(\theta_\ell^{(t)}, \phi_\ell^{(t)}) \, 
e^{-j 2 \pi f \tau_\ell}
$}
\end{equation}
}where $H\in \mathbb{C}^{N_t \times N_r}$ denotes the CSI matrix, with elements $H_{i,j}[f]$ representing the complex channel gain between transmission antenna $i$ to receiver antenna $j$ at frequency $f$. $L$ is the number of propagation paths (including line-of-sight, specular reflection, and scatters), $\alpha_\ell$ and $\tau_\ell$ are the complex gain and propagation delay of the $\ell$-th path, $a_{t,j} \ , \ a_{r,i} $ represent the array responses of the $j$-th transmit and $i$-th receive antennas, and $\theta_\ell^{(t/r)}, \phi_\ell^{(t/r)}$ are the elevation and azimuth angles of departure/arrival.

We can observe that the underlying structure of the wireless channel is primarily determined by the wireless medium and the scene properties as well as physical attributes such as the relative location of the user which together define the propagation paths' properties $(\alpha_\ell ,\tau_\ell,\theta_\ell^{(t/r)}, \phi_\ell^{(t/r)})$ in Eq.~\ref{eq:wireless_channel}. This important observation guides the selection of appropriate modality types to extract physically coherent and generalizable representations within the foundational model. Consequently, we design our foundational model to incorporate three key components: the channel state information (i.e., $H \in \mathbb{C}^{N_t \times N_r}$ in Eq.~\ref{eq:wireless_channel}), the static 3D representation of the wireless environment, and the relative user location. %For simplicity, in this work, we consider a single carrier frequency and treat only the user location as the variable input. %We note that the proposed foundational model can be readily extended to multiple frequencies, which we leave for future work.

%\vspace{-3mm}
\subsection{Foundational Model}
\vspace{-1mm}
\textbf{Input Data Representation:} Given the selected modality types, we represent the CSI associated with each receive antenna as a three-dimensional tensor of size $(N_t(x), N_t(y), 2)$, corresponding to a transmit array with $N_t(x)$ and $N_t(y)$ antennas along the horizontal and vertical dimensions, respectively. The third dimension encodes the real and imaginary components of the complex channel gain. In the current implementation, we assume single-frequency CSI; however, this representation can be readily extended to wideband operation by introducing an additional dimension to capture frequency-dependent channel variations. In this work, we consider a $32 \times 32$ antenna array deployed at the base station at a height of 15~m and operating at a carrier frequency of 28.5~GHz in the millimeter-wave band. The scene properties are represented by a two-dimensional matrix of size $(N \times N)$, where each entry denotes the building height at the corresponding spatial location. We assume a coverage area of $200\,\mathrm{m} \times 200\,\mathrm{m}$ with a spatial resolution of 1~m, with the base station located at the top-center of the scene. Finally, the user location information is represented as a two-dimensional vector encoding the relative position of the user with respect to the base station.

\vspace{-1mm}
\textbf{Backbone Model Architecture:} The overview of the multi-modal foundational model is illustrated in Fig.\ref{fig:main_abstract_img}. 
Inspired by the Vision Transformer (ViT)~\cite{dosovitskiy2020image}, we partition the complex CSI matrix and the 2D static scene property tensor into fixed-size non-overlapping patches, which are subsequently flattened. A modality-specific linear projection is then applied to map the patches from different modalities into a shared token space (see Appendix~\ref{app:backbone_model} for more details). For each modality, the resulting tokens are augmented with learnable positional embeddings to encode the physical location of patches within the antenna array and the wireless environment, as well as modality embeddings to distinguish different input types. On top of the input token sequence, we introduce a learnable $[\mathrm{PHYSC}]$ token that serves as a global aggregator, summarizing physically relevant information across modalities and capturing both intra- and inter-modality correspondences. The encoded $[\mathrm{PHYSC}]$ token is subsequently used as the learned representation for downstream tasks.

%with embedding dimension $d=128$. We adopt squared patch sizes of $4 \times 4$ and $10 \times 10$ for the CSI and 2D scene inputs, respectively. 

To enable early and effective cross-modality interaction, we employ a single-stream Transformer encoder with merged-token self-attention~\cite{vaswani2017attention,geng2022multimodal}, where tokens from different modalities jointly attend to each other from the first layer onward to capture complex inter- and intra-modality dependencies (see Appendix~\ref{app:backbone_model} for more details). 
%The encoder consists of multiple transformer blocks, allowing the model to capture complex inter- and intra-modality dependencies. We use eight transformer layers with eight attention heads per layer. 
During pretraining, latent representations of the visible tokens are concatenated with modality-specific learnable mask tokens and passed to a lightweight transformer decoder. This decoder consists of a small stack of transformer blocks followed by task-specific projection heads that map token embeddings to the outputs required for the reconstruction-based self-supervised objectives. The designed foundational model comprises 2.4 million trainable parameters, orders of magnitude smaller than state-of-the-art VLMs, thereby enabling inference on the order of milliseconds on Base Stations and satisfying the stringent latency requirements of cellular networks.
%\yg{can we have some numbers like order of magnitude time for inference?}

%\vspace{-3mm}
\subsection{Physics-Informed Self-Supervised Pretraining}\label{pretraining}
\vspace{-1mm}
During pretraining, our objective is to extract physically meaningful correspondences across different modalities and leverage the representational capacity of the foundational model to learn global embeddings that can be effectively adapted to diverse downstream applications. Unlike conventional VLMs that rely on contrastive learning objectives (e.g., CLIP~\cite{radford2021learning}), which enforce alignment by pulling positive image--text pairs closer in the embedding space while pushing negative pairs apart, such a formulation is not directly applicable to wireless systems. In particular, different environmental scenes and user locations may result in similar CSI realizations due to the combination of interacting propagation paths. Consequently, defining explicit positive and negative cross-modality pairs between CSI and scene information is ambiguous and potentially misleading.
\newline
To address this challenge, we adopt the masked autoencoder paradigm~\cite{he2022masked,geng2022multimodal} and develop a physics-guided self-supervised pretraining strategy tailored to physical-layer wireless systems.
Specifically, our self-supervised learning framework consists of two complementary components:  
(1) \textit{Reconstruction Loss Minimization}, and  
(2) \textit{Physics-Guided Cross-Modal Distillation}.  

In the first component, we encourage the model to learn physically meaningful associations across modalities by generating multiple viewpoint embeddings that are used to reconstruct masked portions of the input, leveraging inter- and intra-modality dependencies. For the CSI modality, each decoded CSI token is passed through a reconstruction head to predict the complex channel gain corresponding to the associated transmit antenna patch. In contrast, for the scene modality, the model is tasked with reconstructing masked patches of a 2D building-height occupancy grid represented as a binary matrix, rather than performing point-wise reconstruction of detailed 3D scene properties from CSI.
Such fine-grained reconstruction would lead to severe overfitting and conflicts with the objective of learning universal representations. This design choice enforces the extraction of meaningful and generalizable cross-modality relationships grounded in the physics of EM wave propagation, including line-of-sight (LOS) vs Non-LOS propagation, as well as reflection and diffraction properties by predicting whether a patch in a certain position contains obstructing/reflecting or free space properties based on observed CSI structure. 
%The occupancy grid is constructed as a binary matrix, where a patch is labeled as occupied if more than 60\% of its area is covered by buildings.

To ensure physical relevance during reconstruction, we exploit the fact that the wireless channel experienced by a given user is primarily determined by interactions between EM waves and surrounding reflective or obstructive structures. As a result, reconstructing arbitrary or physically irrelevant scene regions that do not influence the channel for a specific user location is neither meaningful nor learnable, and would lead to noisy or non-convergent training behavior. Accordingly, while CSI tokens are randomly masked, scene tokens are selectively masked based on their spatial relevance: only patches located within a predefined distance threshold from the line connecting the base station and the user location are randomly masked during pretraining.

\begin{figure*}[t]
    \centering
\includegraphics[width=0.85\textwidth]{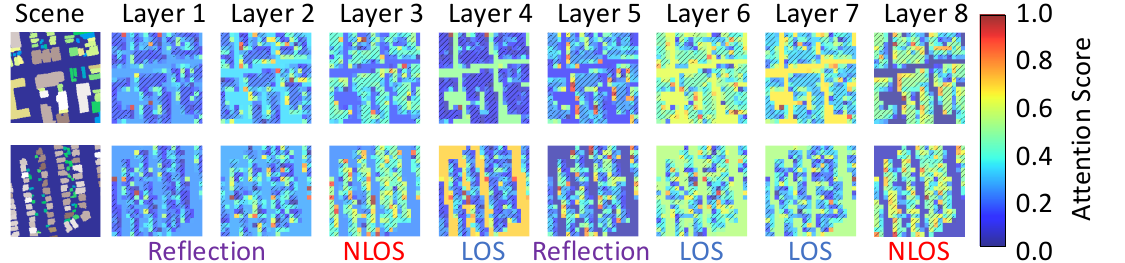} 
\vspace{-3mm} 
\caption{\small $[\mathrm{PHYSC}]$ token attention score on scene patches reveals learned meaningful patterns that reflect EM wave propagation characteristics when interacting with scene components, including LOS/NLOS and reflection effects.}
\vspace{-4mm}
    \label{fig:attention_pattern}
\end{figure*}

\begin{figure}[!b]
\centering
\vspace{-8mm}
\includegraphics[width=0.4\textwidth]{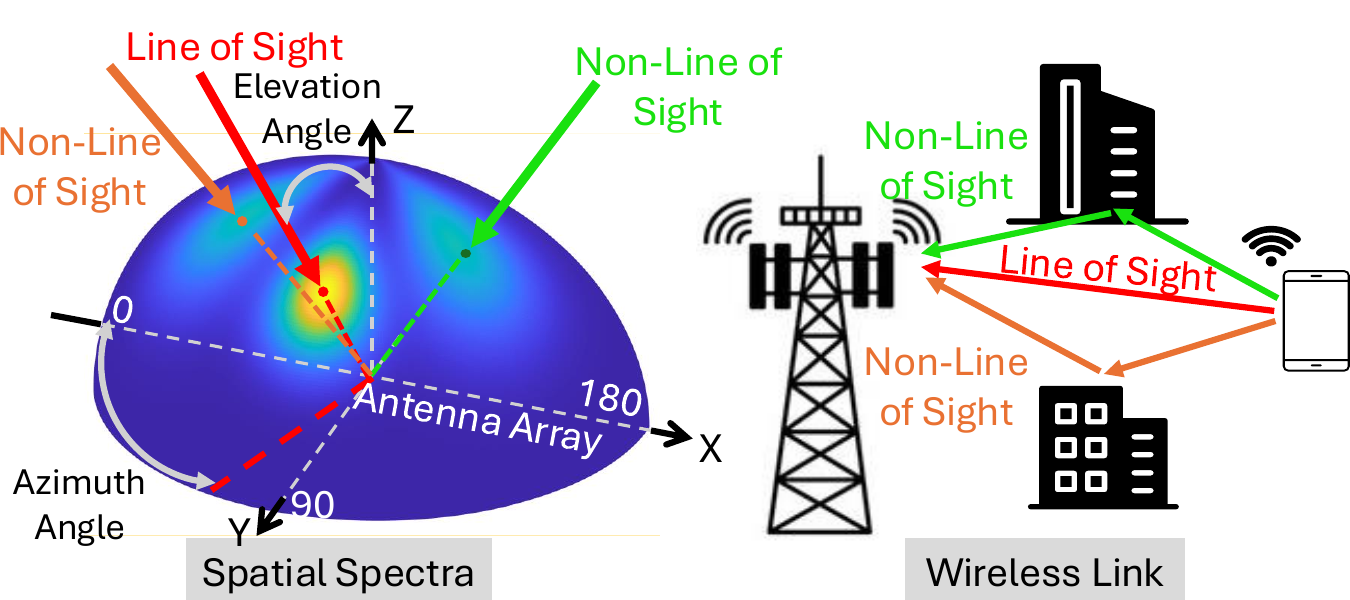} 
\vspace{-4mm}
\caption{\small Illustration of spatial spectrum.}
\vspace{-5mm}
\label{fig:spatial_spectra}
\end{figure}

In the second component, i.e., physics-guided cross-modal distillation, we inject prior physical knowledge that captures the intrinsic associations among modalities into the global aggregator token. Specifically, the model is encouraged to distill physically meaningful information into the encoded $[\mathrm{PHYSC}]$ token such that it can reconstruct a target physical quantity that jointly correlates all modalities. This mechanism promotes robustness under data distribution shifts and generalization to unseen environments by enforcing adherence to physical correspondences among the input modalities, which remain invariant across different settings.
A key question then arises: \textit{Which physical property can effectively capture the shared structural dependencies across all modalities?}  We seek a representation that characterizes how signal energy is distributed over spatial frequencies or directions in a given environment and transmitter/receiver position. \textbf{Spatial spectra} possess this property and are commonly used to reveal dominant propagation directions, such as angles of arrival and departure, which are governed by the wireless propagation medium and can be inferred from the joint observation of all modalities. An illustrative example of spatial spectra and their relationship to the underlying wireless environment is shown in Fig.~\ref{fig:spatial_spectra}. Accordingly, we enforce the reconstruction of a coarse-grained spatial spectrum from the decoded $[\mathrm{PHYSC}]$ token via a linear projection head. This design enables effective physics-guided cross-modal distillation in the encoded token space, anchoring the learned representations to physically interpretable and environment-invariant propagation characteristics.

Finally, the pretraining loss is defined as a weighted sum of four terms: \textit{(i)} masked CSI reconstruction, quantified using the root mean squared error (RMSE) between the estimated and ground-truth complex channel coefficients; \textit{(ii)} masked occupancy-grid reconstruction, formulated as a binary cross-entropy loss between the predicted and true occupancy labels for each scene patch; \textit{(iii)} masked location reconstruction, measured by the RMSE between the estimated and true user position; \textit{(iv) } spatial spectra reconstruction is formulated using a weighted combination of logarithmic- and linear-domain RMSE between the predicted and ground-truth spectra. This design promotes stable training while jointly capturing relative and absolute reconstruction errors, enabling reliable identification of all propagation paths with emphasis on the dominant components. Refer to Appendix~\ref{app:pretraining_loss} for more details on pretraining.  

%\yg{ As I read this  I wonder if you can add a simple graphic that shows spatial spectra and its replationship with EM propagation, environemnt etc}
% We employ a reconstruction-based learning strategy with progressively increasing masking ratios, organized into three high-level reconstruction objectives: 
% (1) predominantly masking CSI tokens, 
% (2) predominantly masking scene information tokens, and 
% (3) moderately masking tokens across all modalities. To facilitate progressive learning and avoid catastrophic forgetting, we interleave easier and harder reconstruction tasks (stage 1 and stage 2), adjusting their sampling probabilities across training epochs. This ensures that the foundational model gradually learns to capture both inter- and intra-modality correlations. In the second stage, location information tokens are masked with a 50\% probability for each data point during moderately masked tasks in stage 2. Table~\ref{tab:masking_strategy} summarizes the masking ratios for each modality.

% \begin{table}[h]
% \centering
% \caption{Masking strategy across stages}
% \label{tab:masking_strategy}
% \scriptsize
% \resizebox{\columnwidth}{!}{%
% \begin{tabular}{|c|c|c|c|c|c|}
% \hline
%  & CSI Large & Heigth Large & CSI Mod & Height Mod\\
% \hline
% Stage 1 & 0.5 & 0.1 & 0.3 & 0.05 \\
% \hline
% Stage 2 & 0.7-0.9 & 0.2-0.3 & 0.5-0.7 & 0.1-0.2  \\
% \hline
% \end{tabular}
% }
% \end{table} 
\vspace{-3mm}
\subsection{Pretraining Dataset}
\label{dataset}
\vspace{-2mm}

We construct the pretraining dataset by simulating EM wave propagation using the Sionna Ray-Tracing~\cite{hoydis2023sionna} across approximately 10{,}000 distinct outdoor urban environments spanning more than 200 cities worldwide. Each environment represents a $200\,\mathrm{m}$ square coverage area that serves as the wireless propagation medium. For each scene, we extract detailed 3D environmental information using OpenStreetMap~\cite{OSM} data and the Geo2SigMap~\cite{li2024geo2sigmap} processing pipeline, which provides building geometry and height information. Ray-tracing simulations are then performed at a carrier frequency of 28.5~GHz in the millimeter-wave band for a base station equipped with a uniform rectangular transmit antenna array. For each scene, wireless channels are generated for 300-500 users randomly distributed within each coverage area. 
The resulting dataset comprises over 3.5 million data samples, each consisting of a complex-valued channel tensor, the corresponding user location, and a 2D matrix representing the building-height map of the scene. We note that this is the first large-scale multi-modal wireless channel dataset in mm-Wave frequency band that spans across 10{,}000 distinct realistic outdoor scene information. 
%We utilize 80\% of the generated dataset for pretraining the foundational model. 
To facilitate reproducibility and further research, we will publicly release both the dataset and the trained model upon publication.

%\vspace{-3mm}
\subsection{Interpreting Attention Patterns in the Pretrained Foundational Model Through a Physical Lens}
\vspace{-1mm}
We use the self-supervised learning framework introduced in Section~\ref{pretraining} together with the dataset described in Section~\ref{dataset} to pretrain the foundational model for 50 epochs, utilizing four NVIDIA A100 80GB PCIe GPUs. The multi-modal reconstruction capability of the pretrained model on unseen samples is detailed in Appendix~\ref{app:pretraining_objectives}.  

%\vspace{-2mm}
Beyond reconstruction, we focus on interpreting the attention patterns of the pretrained foundational model through a physics-informed lens. Specifically, we analyze the attention scores of the $[\mathrm{PHYSC}]$ token over scene geometry tokens across different transformer layers to understand the patterns that contribute to its encoded universal representation. Fig.~\ref{fig:attention_pattern} visualizes the normalized attention scores of the $[\mathrm{PHYSC}]$ token over height-map tokens, averaged across attention heads for each layer, for two unseen wireless environments. Interestingly, we observe that the physics-guided self-supervised pretraining enables the foundational model to capture physically meaningful attention patterns. The $[\mathrm{PHYSC}]$ token selectively attends to scene regions corresponding to key EM wave propagation phenomena: \textit{(i)} LOS/NLOS conditions, where a higher attention score is placed on free-space/obstructing regions in specific layers, and \textit{(ii)} reflection and diffraction effects, where attention is concentrated on scene tokens representing building edges that contribute to specular reflections and diffractions. These patterns illustrate that the model forms its universal representation by emphasizing physically relevant features in forming the wireless channel. 

\section{Evaluation of Pretrained Foundational Model to Downstream Applications}
\vspace{-1mm}
In this section, we leverage the task-agnostic pretrained foundational model for a range of wireless downstream tasks to evaluate how effectively the model’s learned universal understanding of the wireless channel, captured by the $[\mathrm{PHYSC}]$ token, can be transferred to new applications with limited labeled data. To this end, we define several AI-driven physical-layer tasks and construct corresponding datasets to systematically evaluate model generalization.
%\yg{you can shorten this par if you are out of space}
\vspace{-2mm}
\subsection{Down Stream Tasks} 
\vspace{-1mm}
\textbf{Wireless User Localization:}
Accurate localization of user equipment in wireless networks is a critical capability for a wide range of applications, including beamforming, resource allocation, and context-aware services~\cite{trevlakis2023localization}.
In complex wireless environments, multipath propagation induces distinct signatures in CSI that depend on the relative position of the user equipment with respect to the base station, providing valuable cues for localization. However, these signatures are strongly influenced by the underlying 3D scene geometry and wireless medium, making accurate identification of the scene distribution essential for robust localization. In this application, we evaluate the ability of the proposed foundational model to learn transferable embeddings that enable accurate localization from partial CSI with limited labeled on-site data. Specifically, we assume that the scene information and 25\% of the CSI are available to the model, while the location is masked.
%\yg{ I thought you don't have the full 3D scene knowledge. Then this is an easy task, no? brute force would do it. }

We consider two standard localization paradigms:
\vspace{-1mm}

\textbf{(1) Scene-specific deployment.}  
In the first setting, we assume a dedicated localization model for each deployment environment. Models are trained using a subset of labeled data collected from a given scene and evaluated on the remaining samples to assess within-scene generalization and data collection requirements for robust performance. For this task, we apply linear probing on both the encoded $[\mathrm{PHYSC}]$ token from our model and the raw CSI used as the baseline. The dataset for this experiment comprises four deployment areas (see Appendix~\ref{app:localization_setting}), each containing 10{,}000 CSI--location labeled data points.

\vspace{-1mm}
\textbf{(2) Cross-scene generalization.}  
In the second setting, we consider a single localization model shared across multiple deployment scenarios. The model is trained on a large-scale multi-scene dataset and evaluated on previously unseen environments under zero-shot and few-shot adaptation regimes, where limited labeled data from the target environment are available for fine-tuning. 
Localization is performed using a lightweight MLP operating on the foundational model's encoded $[\mathrm{PHYSC}]$ token, and a CNN applied to the available raw CSI that serves as the task-specific baseline. For training, we construct a task-specific dataset consisting of approximately 345{,}000 labeled samples spanning 830 distinct scenes. The trained models are then evaluated on four unseen deployment environments (see Appendix~\ref{app:localization_setting}), each with a total of 2{,}000 labeled data points collected for testing and adaptation. To reduce overfitting for the baseline and ensure a fair comparison, both models are fine-tuned using a smaller learning rate than in the scene-specific setting where models are trained from scratch (see Appendix~\ref{app:localization_setting}).

\begin{figure*}[t]
    \centering
\includegraphics[width=1\textwidth]{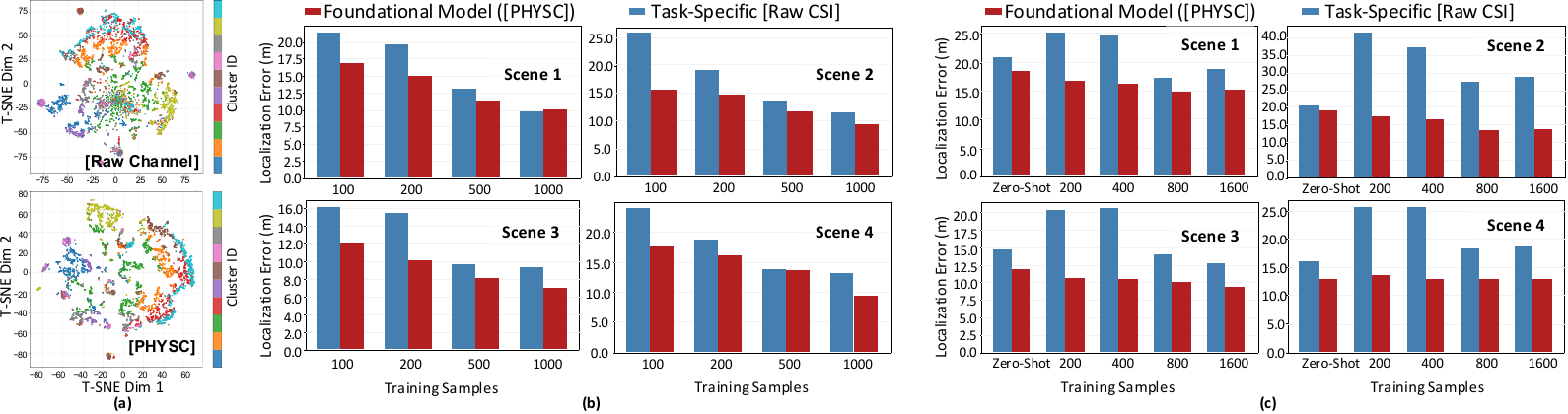} 
\vspace{-8mm} 
\caption{\small Results on adapting the foundational model to wireless localization: (a) t-SNE visualization of raw CSI and the encoded $[\mathrm{PHYSC}]$ token; localization error vs the number of training samples in (b) scene-specific, and (c) cross-scene generalization settings.}
\vspace{-5mm}
    \label{fig:Localization}
\end{figure*}

\textbf{MIMO Hybrid Precoding:}  
MIMO precoding is a signal processing technique used in multi-antenna wireless systems to spatially direct transmitted signals toward intended receivers, thereby enabling multi-stream concurrent transmission and improving the overall system throughput. Correctly identifying the MIMO precoding, especially in wireless systems with hybrid analog-digital architectures~\cite{molisch2017hybrid}, is critical for optimizing the sum-rate. In hybrid MIMO scenarios, the non-convex optimization problem of finding the optimal precoding matrix is conventionally divided into two steps. First, the analog beamforming/combining vectors are selected from a predefined quantized codebook via exhaustive search to maximize the sum of the perceived signal-to-noise ratio across all users. Second, the optimal baseband precoders/combiners are calculated analytically using the effective channel~\cite{alkhateeb2016frequency} (see detailed theoretical derivations in Appendix~\ref{app:mimo}). However, exhaustive search over the entire codebook entails a prohibitive wireless transmission overhead~\cite{8871119}. To overcome this, prior work has shown promise of AI-driven solutions to approach this classification problem with partial observations~\cite{li2019deep}. In this paper, we formulate two variants of the AI-enabled hybrid MIMO precoding as downstream tasks to evaluate the generalizability of the foundational model on critical wireless communication tasks.

%Although extensive theoretical work has derived optimal precoding weights based on full CSI, such approaches are impractical in large-scale MIMO systems due to the prohibitive overhead of full CSI acquisition~\cite{8871119}.  

\textbf{(1) Multi-User MIMO.}  
In the first task, we consider a multi-user MIMO setting in which a base station equipped with $8$ Radio Frequency (RF) chains serves a cluster of $8$ single-antenna users, each with one RF chain. The problem is formulated as identifying the optimal analog beamforming index for each user from partial CSI (5\% available), which can then be mapped to the full hybrid precoding matrix. We perform linear probing using the encoded $[\mathrm{PHYSC}]$ token for the foundational model and raw partial CSI of comparable dimensionality for the baseline, to classify the beam index for each channel realization. The beam index is selected from a codebook of size $1024$, corresponding to a $32 \times 32$ antenna array. We further use exhaustive search over the codebook as a local upper bound to find the optimal beam label and benchmark the performance of all models in terms of a practical wireless system KPI, namely the achieved sum-rate (see Appendix~\ref{app:mu_hybrid}). For this task, we collect a dataset comprising 27{,}000 CSI samples across 61 distinct scenes.

\begin{figure*}[t]
    \centering
\vspace{-2mm}
\includegraphics[width=1\textwidth]{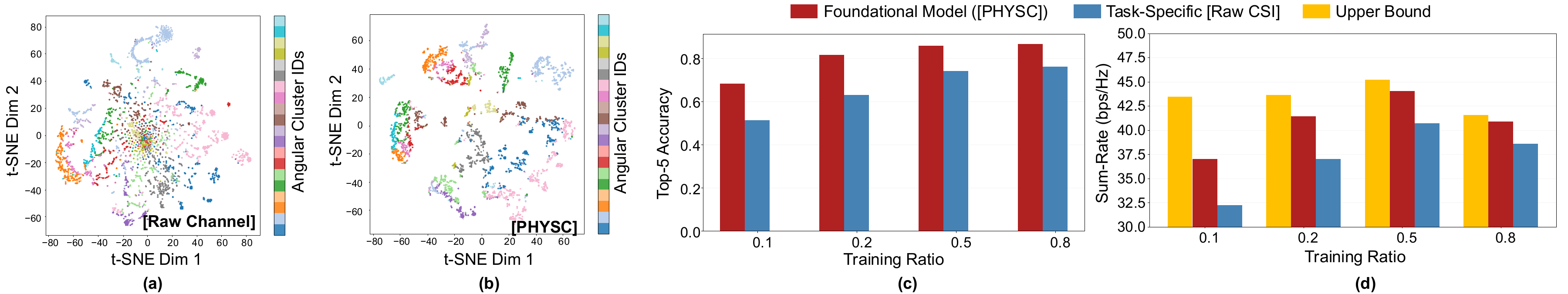} 
\vspace{-9mm} 
\caption{\small Results on adapting foundational model to MU-MIMO: t-SNE visualization of (a) raw channel and (b) encoded $[\mathrm{PHYSC}]$ token; (c): top-5 classification accuracy of detecting optimal beam index, and (d): achieved sum-rate as a function of training ratio.}
\vspace{-5mm}
    \label{fig:MIMO_Results}
\end{figure*}

\vspace{-1mm}
\textbf{(2) Single-User MIMO.}  
In the second task, we consider a single-user MIMO setting in which a base station equipped with a $32 \times 32$ antenna array with two RF chains serves a user equipped with four antennas and two RF chains. The task is formulated as identifying the top-2 optimal analog beamforming and combining indices from partial CSI (5\% available). These indices are subsequently mapped to the full hybrid precoding matrices at the transmitter and receiver. The two transmit beam indices are selected from a codebook of size $N_{\mathrm{tx}}=1024$, while the two receive beam indices are selected from a codebook of size $N_{\mathrm{rx}} = 4$. 
%We select this setting to evaluate the adaptability of the foundational model to previously unseen system structures. 
We employ two separate linear-probing classifiers to independently predict the transmitter and receiver beam indices from a given feature vector. The feature vector is formed using either the concatenated encoded $[\mathrm{PHYSC}]$ tokens from the foundational model across all receiver antennas or the raw CSI–receiver pairs as the baseline. For this task, we construct a dataset comprising 34{,}000 CSI samples of dimension $32 \times 32 \times 4$ collected across 61 distinct scenes. Importantly, the structure of this task differs fundamentally from the pretraining phase, where receivers are assumed to be single-antenna and receiver-side spatial dimensions are not modeled. Consequently, fine-tuning the base model is required to handle multi-antenna receivers. A more general framework would explicitly model receiver dimensionality during pretraining, which we leave as future work.

The optimal beamforming labels are obtained via exhaustive search over all codebook beam index combinations and used to compute the local upper-bound sum-rate for benchmarking models' performance (see Appendix~\ref{app:su_hybrid}).

%\vspace{-3mm}
\subsection{Results and Key Takeaways}
\vspace{-1mm}
%In this section, we provide a systematic evaluation study on important wireless applications to demonstrate the strong capacities of the foundational model to adapt to new tasks.

\textbf{Localization.} To examine how the foundational model’s learned representations support localization, we first visualize the t-SNE~\cite{maaten2008visualizing} projections of the encoded $[\mathrm{PHYSC}]$ token and raw CSI in Fig.~\ref{fig:Localization}(a). Data points are colored according to representative cluster IDs that group users based on their positional proximity. The embeddings produced by the foundational model exhibit well-separated and distinguishable clusters, in contrast to the raw CSI features. This separation indicates that the foundational model captures intrinsic wireless channel characteristics in a semantically meaningful latent space, yielding representations that are more suitable for localization tasks.

Fig.~\ref{fig:Localization}(b) reports the median localization error under a 10dB SNR condition across four different scenes for scene-specific deployment, comparing the foundational model embeddings against raw CSI features as a function of the number of training samples. The results demonstrate a consistent performance gain when leveraging the foundational model, particularly in data-limited regimes. Specifically, the foundational representation improves localization accuracy by approximately 5–10 meters, corresponding to more than a 20\% relative reduction in error compared to the baseline.

Finally, Fig.~\ref{fig:Localization}(c) evaluates localization performance under a cross-scene generalization setting. The foundational model consistently achieves improved performance in zero-shot adaptation to previously unseen environments, owing to its generalizable representations that encode underlying scene geometry and its interaction with wireless propagation. Furthermore, as additional labeled samples become available during fine-tuning, the foundational model continues to reduce its inference error. In contrast, the baseline relying on raw CSI degrades in performance when fine-tuned with only a few labeled samples. 
This behavior highlights the advantage of the foundational model in capturing transferable structural properties of the environment and adapting to new scene distribution, while the task-specific model using raw CSI exhibits limited robustness to distribution shifts, leading to degraded localization accuracy in unseen regions. 

%\yg{see slack about raw channel. add scene info to appendix for all plots}

\begin{figure}[!b]
\centering
\vspace{-4mm}
\includegraphics[width=0.45\textwidth]{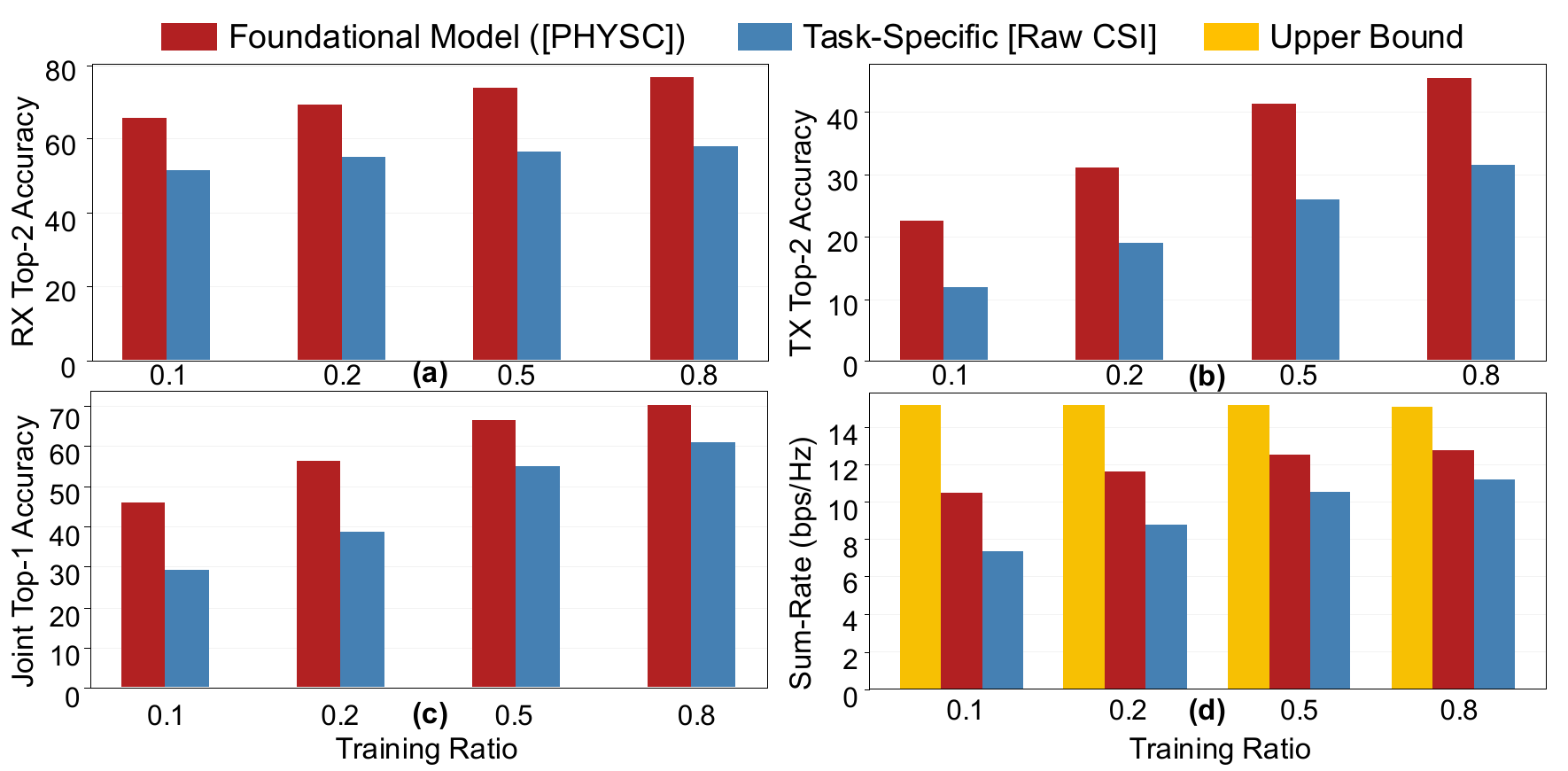} 
\vspace{-4mm}
\caption{\small Single-user MIMO communication: top-2 accuracy for (a) receiver and (b) transmitter beam prediction; (c) joint Tx–Rx dominant beam detection; and (d) achieved sum-rate.}
\vspace{-5mm}
\label{fig:Single_MIMO_Results}
\end{figure}

\textbf{MIMO Hybrid Precoding.}
Next, we examine the adaptability of the foundational model’s learned representations to MIMO hybrid precoding tasks. We visualize the t-SNE projections of the raw CSI and the encoded $[\mathrm{PHYSC}]$ token in Fig.~\ref{fig:MIMO_Results}(a)–(b), where data points are colored according to angular clusters IDs corresponding to groups of channel conditions associated with optimal beam indices that exhibit spatial proximity in angular direction. The encoded $[\mathrm{PHYSC}]$ token demonstrates significantly improved class separation compared to raw CSI, indicating enhanced interpretability and suitability for hybrid precoding decisions.

Figs.~\ref{fig:MIMO_Results}(c)–(d) report the top-5 beam detection accuracy and the achieved sum-rate for the foundational model, the task-specific baseline, and an upper bound obtained via exhaustive search, as functions of the training data ratio. The results illustrate the strong capability of the foundational model to adapt to hybrid MU-MIMO precoding, achieving consistently higher beam detection accuracy and sum-rate performance, particularly in data-limited regimes. Notably, the foundational model attains near-optimal sum-rate performance using only 20\% of the available training samples and improves beam detection accuracy by nearly 20\% compared to the task-specific baseline. Further details, including the empirical cumulative distribution function of achieved sum-rate across all user clusters and an ablation study evaluating the impact of different classification head configurations on model performance, are provided in Appendix~\ref{app:MU_MIMO_Imp}.

In addition, we also evaluate the performance of a single-user MIMO precoding system. Fig.~\ref{fig:Single_MIMO_Results} summarizes the results by comparing the performance of the fine-tuned foundational model with the task-specific baseline with different training ratios. Figures~\ref{fig:Single_MIMO_Results}(a) and (b) show the accuracy of identifying the two optimal beams at the receiver and transmitter, respectively. Fig.~\ref{fig:Single_MIMO_Results}(c) reports the accuracy of detecting the joint dominant transmitter--receiver beam pair, and Fig.~\ref{fig:Single_MIMO_Results}(d) presents the corresponding achieved sum-rate. These results demonstrate the capability of the foundational model to adapt to new problem structures through appropriate fine-tuning, leading to enhanced wireless system performance. In particular, the foundational model provides higher beam detection accuracy, especially for the more challenging components of identifying transmitter beam indices and joint dominant beam pairs, where the number of possible combinations is much larger, achieving approximately a 20\% accuracy improvement in data-limited scenarios. Further, it improves the achieved sum-rate by nearly 42\% when trained with only 10\% of the available data. Notably, this task involves a more challenging optimization problem, which, even under full CSI knowledge, lacks a closed-form solution and requires a computationally intensive exhaustive search to obtain the upper bound performance (see Appendix ~\ref{app:su_hybrid}). Therefore, achieving near-optimal performance, especially under data-limited conditions, is an inherently complex problem.

%\yg{see slack- add sentences to clarify the 3 concerns i raised}

%\vspace{-5mm}
\section{Discussion}
This paper introduces a first-of-its-kind task-agnostic, multimodal foundational model for wireless systems, designed to learn generalizable representations grounded in the physics of EM wave propagation. We propose a self-supervised pretraining framework that explicitly incorporates physical priors into the learning paradigm through physics-guided cross-modal distillation. We introduce a dedicated $[\mathrm{PHYSC}]$ token that serves as a global aggregator across input modalities to encode and propagate physical knowledge of EM wave interactions with scene components by guiding the model to learn cross-modal associations that optimally reconstruct \textit{the spatial spectra}. %Our design enables the model to internalize fundamental EM characteristics and effectively transfer this knowledge across diverse wireless applications. 
%To facilitate large-scale pretraining and systematic evaluation, we develop a multimodal wireless dataset, together with standardized downstream task benchmarks.
Rigorous evaluations demonstrate that our design enables the model to internalize EM propagation characteristics and effectively transfer this knowledge across wireless applications. The proposed foundational model improves localization accuracy by 20\% and communication link quality by up to 40\% over task specific baseline, highlighting the model's effectiveness in data-constrained settings and its potential as a backbone for cross-task generalization in future wireless networks.

\newpage

\bibliography{references}
\bibliographystyle{icml2025}

%%%%%%%%%%%%%%%%%%%%%%%%%%%%%%%%%%%%%%%%%%%%%%%%%%%%%%%%%%%%%%%%%%%%%%%%%%%%%%%
%%%%%%%%%%%%%%%%%%%%%%%%%%%%%%%%%%%%%%%%%%%%%%%%%%%%%%%%%%%%%%%%%%%%%%%%%%%%%%%
% APPENDIX
%%%%%%%%%%%%%%%%%%%%%%%%%%%%%%%%%%%%%%%%%%%%%%%%%%%%%%%%%%%%%%%%%%%%%%%%%%%%%%%
%%%%%%%%%%%%%%%%%%%%%%%%%%%%%%%%%%%%%%%%%%%%%%%%%%%%%%%%%%%%%%%%%%%%%%%%%%%%%%%
\newpage
\appendix
\onecolumn

\section{Details on Backbone Model Architecture}
\label{app:backbone_model}
We adopt a multi-modal masked autoencoder architecture with modality-specific tokenizers based on non-overlapping 2D patch embeddings. Complex CSI is represented using two channels (real and imaginary) over a $32 \times 32$ grid and tokenized with a patch size of $4 \times 4$, yielding 64 CSI tokens, while the height/occupancy map is processed as a single-channel $200 \times 200$ image with $10 \times 10$ patches, producing 400 height tokens. Each modality uses a convolutional patch projection followed by learnable positional embeddings and modality embeddings. Receiver location is embedded via a linear projection from 2D coordinates to the model dimension, and a learnable $[\mathrm{PHYSC}]$ token is introduced to aggregate global physical information. Separator tokens are inserted between modalities to preserve structural boundaries. All tokens are projected to a shared embedding dimension of 128.

The encoder consists of 8 Transformer layers with 4 attention heads and an MLP expansion ratio of 4, while the decoder uses 2 layers with the same embedding size and number of heads. Both encoder and decoder employ standard pre-norm multi-head self-attention and GELU-activated feed-forward blocks. Reconstruction is performed via modality-specific projection heads: masked CSI patches are recovered using a linear head mapping to $2 \times 4^2$ values per patch, masked receiver locations are regressed through a linear 2D head, and masked height tokens are mapped to occupancy logits using a scalar linear head. The $[\mathrm{PHYSC}]$ token is decoded through a lightweight convolutional projection head that reshapes the embedding into a coarse spatial feature map (32 channels) and applies convolutional refinement to produce a single-channel physical heatmap. Learnable mask tokens are used for each modality, and decoder-side positional and modality embeddings are added prior to reconstruction. This design enables joint representation learning across CSI, geometry, and localization while supporting masked reconstruction and downstream adaptation.

\section{Pretraining with Physics Informed Self-Supervised Learning}
\label{app:pretraining_loss}

In this part, we provide detail formulation of the pretraining objectives used in our physics-guided self-supervised learning framework. The overall pretraining loss consists of four components: (i) masked CSI reconstruction, (ii) masked location reconstruction, (iii) masked occupancy-grid reconstruction, and (iv) physics-informed spatial spectra reconstruction.

\subsection{Masked CSI Reconstruction Loss}
Let $\mathbf{H} \in \mathbb{C}^{N_t}$ denote the ground-truth complex channel state information (CSI) matrix between the transmitter with $N_t$ antennas and a single antenna receiver. During pretraining, a subset of CSI patches indexed by the set $\mathcal{M}_{\mathrm{CSI}}$ is masked and reconstructed by the model, yielding the estimate $\hat{\mathbf{H}}$.

The CSI reconstruction loss is defined as the root mean squared error (RMSE) over the masked CSI entries:
\begin{equation}
\mathcal{L}_{\mathrm{CSI}}
=
\sqrt{
\frac{1}{|\mathcal{M}_{\mathrm{CSI}}|}
\sum_{(i,j)\in\mathcal{M}_{\mathrm{CSI}}}
\Big(
\big( |\hat{H}_{i,j}| - |H_{i,j}| \big)^2
+
\big( \angle \hat{H}_{i,j} - \angle H_{i,j} \big)^2
\Big)
}.
\label{eq:csi_rmse}
\end{equation}

This loss enforces accurate reconstruction of both the magnitude and phase of the complex-valued channel coefficients from partially observed CSI and scene information, thereby encouraging the model to learn embeddings that capture the impact of the underlying wireless propagation medium and local CSI structure on the resulting channel behavior.

\subsection{Masked Location Reconstruction Loss}

Let $\mathbf{p} = [x, y]^T \in \mathbb{R}^2$ denote the true user location and $\hat{\mathbf{p}}$. During pretraining, the location token is randomly masked in those isntances the model is forced to estimate the user position  
from the decoded token corresponding to this physical attribute. The location reconstruction loss is defined as
\begin{equation}
\mathcal{L}_{\mathrm{loc}}
=
\sqrt{
\left\|
\hat{\mathbf{p}} - \mathbf{p}
\right\|_2^2
}.
\label{eq:location_rmse}
\end{equation}

This term encourages the model to embed geometry-aware information and multimodal wireless channel associations, enabling it to capture the underlying signatures induced by electromagnetic wave propagation and to extract physically meaningful features that are informative for user positioning.

\subsubsection{Masked Occupancy-Grid Reconstruction Loss}

Let $\mathbf{O} \in \{0,1\}^{N_s}$ denote the binary occupancy grid of the environment, where $O_k = 1$ indicates that the $k$-th scene patch is occupied (i.e., more than $60\%$ covered by buildings), and $O_k = 0$ otherwise. Let $\hat{O}_k \in [0,1]$ be the predicted occupancy probability for patch $k$. The reconstruction is performed only over masked and spatially relevant scene patches indexed by $\mathcal{M}_{\mathrm{occ}}$. To ensure that the reconstruction loss focuses on physically meaningful regions, the set of masked scene patches $\mathcal{M}_{\mathrm{occ}}$ is selected based on their spatial relevance to the wireless link. Specifically, scene patches are prioritized according to their perpendicular distance to the line segment connecting the transmitter and receiver locations. Patches located closer to the TX--RX line are more likely to be masked, as these regions are more likely to interact with propagating electromagnetic waves. The masking region is adaptively expanded until a fixed masking ratio is satisfied, ensuring a consistent number of masked patches while preserving physical relevance. 
The occupancy reconstruction loss is defined using binary cross-entropy (BCE):
\begin{equation}
\mathcal{L}_{\mathrm{occ}}
=
-\frac{1}{|\mathcal{M}_{\mathrm{occ}}|}
\sum_{k\in\mathcal{M}_{\mathrm{occ}}}
\left[
O_k \log(\hat{O}_k)
+
(1-O_k)\log(1-\hat{O}_k)
\right].
\label{eq:occ_bce}
\end{equation}

This loss captures coarse-grained structural information of the environment that is physically relevant to electromagnetic wave propagation by enforcing the model to attend to CSI tokens in a manner that extracts the impact of various wave interactions (e.g., LOS/NLOS conditions, reflection, and diffraction) on the wireless channel.

\subsubsection{Spatial Spectrum Construction}

To inject prior physical knowledge into the global aggregator token, we enforce the reconstruction of a coarse-grained spatial spectrum that captures dominant propagation directions.

Given the full CSI matrix $\mathbf{H}$, the spatial spectrum is computed via a discrete Fourier transform (DFT) across the antenna domain. For a uniform linear array (ULA), the spatial spectrum $\mathbf{S} \in \mathbb{R}^{N_t}$ is defined as
\begin{equation}
\mathbf{S}
=
\left|
\mathbf{F}_{N_t}
\mathbf{h}
\right|^2,
\label{eq:spatial_spectrum}
\end{equation}
where $\mathbf{h}$ denotes the vectorized CSI across the transmit antennas, and $\mathbf{F}_{N_t}$ is the $N_t$-point DFT matrix. The magnitude-squared operation yields a power distribution over spatial frequencies (or angles).

Let $\mathbf{S}$ and $\hat{\mathbf{S}}$ denote the ground-truth and reconstructed spatial spectra, respectively, where $\hat{\mathbf{S}}$ is inferred from partial CSI through the encoded $[\mathrm{PHYSC}]$ token. To ensure numerical stability and balanced sensitivity to both absolute and relative errors, we define the spatial spectrum loss as a weighted combination of linear- and logarithmic-domain RMSEs:
\begin{equation}
\mathcal{L}_{\mathrm{spec}}
=
\alpha
\sqrt{
\frac{1}{N_t}
\left\|
\hat{\mathbf{S}} - \mathbf{S}
\right\|_2^2
}
+
(1-\alpha)
\sqrt{
\frac{1}{N_t}
\left\|
\log(\hat{\mathbf{S}}+\epsilon)
-
\log(\mathbf{S}+\epsilon)
\right\|_2^2
},
\label{eq:spectrum_loss}
\end{equation}
where $\alpha \in [0,1]$ controls the trade-off between linear and logarithmic errors, and $\epsilon$ is a small constant to avoid numerical instability.

\subsection{Overall Pretraining Objective}

The final pretraining loss is defined as a weighted sum of all components:
\begin{equation}
\mathcal{L}_{\mathrm{total}}
=
\lambda_{\mathrm{CSI}} \mathcal{L}_{\mathrm{CSI}}
+
\lambda_{\mathrm{loc}} \mathcal{L}_{\mathrm{loc}}
+
\lambda_{\mathrm{occ}} \mathcal{L}_{\mathrm{occ}}
+
\lambda_{\mathrm{spec}} \mathcal{L}_{\mathrm{spec}},
\label{eq:total_loss}
\end{equation}
where $\lambda_{\mathrm{CSI}}$, $\lambda_{\mathrm{loc}}$, $\lambda_{\mathrm{occ}}$, and $\lambda_{\mathrm{spec}}$ are scalar weighting coefficients that balance the contribution of each loss term.

This formulation enables the model to learn physically meaningful and transferable representations by jointly reconstructing signal-level, geometric, and physics-informed spectral properties from partially observed multimodal inputs.

\subsection{Pretraining Process}
We adopt a reconstruction-based learning strategy with progressively increasing masking ratios, structured around three high-level reconstruction objectives: 
(1) primarily masking CSI tokens, 
(2) primarily masking scene information tokens, and 
(3) moderately masking tokens across all modalities. 
This staged masking approach enables the model to first focus on simpler reconstruction tasks before gradually handling more challenging scenarios. To facilitate progressive learning and mitigate catastrophic forgetting, we interleave easier and harder reconstruction tasks (corresponding to stage 1 and stage 2) and dynamically adjust their sampling probabilities over the course of training (based on training epochs). 
Such a design encourages the foundational model to gradually learn both inter-modality correlations (e.g., between CSI and scene structure) and intra-modality dependencies (e.g., within CSI tokens or scene patches). 
In the second stage, location information tokens are masked with a 50\% probability for each data point during the moderately masked task. 
Table~\ref{tab:masking_strategy} summarizes the masking ratios applied to each modality across different stages.

\begin{table}[h]
\centering
\caption{Masking strategy across stages}
\label{tab:masking_strategy}
\footnotesize
\setlength{\tabcolsep}{8pt} % 
\begin{tabular}{|c|c|c|c|c|}
\hline
 & CSI Large & Height Large & CSI Mod & Height Mod \\
\hline
Stage 1 & 0.5 & 0.1 & 0.3 & 0.05 \\
\hline
Stage 2 & 0.7--0.9 & 0.2--0.3 & 0.5--0.7 & 0.1--0.2 \\
\hline
\end{tabular}
\end{table}

\section{Hybrid MIMO Precoding Formulation}
\label{app:mimo}
\subsection{Multi-User Hybrid MIMO}
\label{app:mu_hybrid}
We consider a downlink multi-user multiple-input multiple-output (MU-MIMO) system in which a base station (BS) equipped with $N_t$ transmit antennas simultaneously serves $K$ single-antenna users over the same time--frequency resource, as illustrated in Figure~\ref{fig:MIMO_Setting} (a). Let $\mathcal{K}=\{1,\dots,K\}$ denote the set of users. The narrowband flat-fading channel between the BS and user $k$ is represented by the vector $\mathbf{h}_k \in \mathbb{C}^{N_t \times 1}$. By stacking the individual user channels, the overall channel matrix is given by
\begin{equation}
\mathbf{H} = 
\begin{bmatrix}
\mathbf{h}_1^{\mathsf{H}} \\
\mathbf{h}_2^{\mathsf{H}} \\
\vdots \\
\mathbf{h}_K^{\mathsf{H}}
\end{bmatrix}
\in \mathbb{C}^{K \times N_t}.
\end{equation}

The BS transmits a symbol vector $\mathbf{s} \in \mathbb{C}^{K \times 1}$ with $\mathbb{E}[\mathbf{s}\mathbf{s}^{\mathsf{H}}]=\mathbf{I}_K$ using a hybrid precoding architecture. The transmitted signal is
\begin{equation}
\mathbf{x} = \mathbf{F}_{\mathrm{RF}} \mathbf{F}_{\mathrm{BB}} \mathbf{s},
\end{equation}
where $\mathbf{F}_{\mathrm{RF}} \in \mathbb{C}^{N_t \times K}$ is the analog (RF) precoder and $\mathbf{F}_{\mathrm{BB}} \in \mathbb{C}^{K \times K}$ is the digital baseband precoder. The received signal at user $k$ is
\begin{equation}
y_k = \mathbf{h}_k^{\mathsf{H}} \mathbf{F}_{\mathrm{RF}} \mathbf{F}_{\mathrm{BB}} \mathbf{s} + n_k,
\end{equation}
where $n_k \sim \mathcal{CN}(0,\sigma^2)$ denotes additive white Gaussian noise.

\subsubsection{Sum-Rate Maximization Problem}

The objective is to design the hybrid precoder $\{\mathbf{F}_{\mathrm{RF}}, \mathbf{F}_{\mathrm{BB}}\}$ to maximize the achievable downlink sum rate under a total transmit power constraint. The resulting optimization problem can be written as
\begin{equation}
\begin{aligned}
\max_{\mathbf{F}_{\mathrm{RF}},\,\mathbf{F}_{\mathrm{BB}}} \quad 
& \sum_{k=1}^{K} 
\log_2\!\left(1 + \mathrm{SINR}_k \right) \\
\text{s.t.} \quad 
& \|\mathbf{F}_{\mathrm{RF}} \mathbf{F}_{\mathrm{BB}}\|_F^2 \le P, \\
& \mathbf{F}_{\mathrm{RF}} \in \mathcal{F},
\end{aligned}
\end{equation}
where $P$ is the transmit power budget, $\mathcal{F}$ denotes the feasible set of analog precoders (codebook) imposed by hardware constraints (e.g., constant-modulus phase shifters), and
\begin{equation}
\mathrm{SINR}_k =
\frac{\left|\mathbf{h}_k^{\mathsf{H}} \mathbf{F}_{\mathrm{RF}} \mathbf{f}_{\mathrm{BB},k}\right|^2}
{\sum_{j\neq k} \left|\mathbf{h}_k^{\mathsf{H}} \mathbf{F}_{\mathrm{RF}} \mathbf{f}_{\mathrm{BB},j}\right|^2 + \sigma^2},
\end{equation}
with $\mathbf{f}_{\mathrm{BB},k}$ denoting the $k$-th column of $\mathbf{F}_{\mathrm{BB}}$.
Due to the non-convex coupling between the analog and digital precoders, we adopt a two-step design approach. It should be noted that this appraoch provide a locally optimal solution and it is not a global optimum. 

\subsubsection{Step 1: Analog Beam Selection from a Codebook}

We assume the BS is equipped with a predefined analog beam codebook containing $N_c$ orthogonal beams
\begin{equation}
\mathcal{C} = \{\mathbf{a}_1, \mathbf{a}_2, \dots, \mathbf{a}_{N_c}\},
\end{equation}
where each codeword $\mathbf{a}_j \in \mathbb{C}^{N_t \times 1}$ satisfies the constant-modulus constraint. The codebook is constructed using a discrete Fourier transform (DFT) representation of a uniform planar array (UPA) with $N_x$ and $N_y$ antenna elements along the $x$- and $y$-axes, respectively, such that $N_t = N_x N_y$ and $N_c = N_x N_y$.

Each codebook index $j$ is uniquely associated with a pair of spatial frequency indices $(k_x,k_y)$ according to a one-to-one mapping
\begin{equation}
j = k_x N_y + k_y,
\qquad
k_x \in \{0,\dots,N_x-1\}, \;
k_y \in \{0,\dots,N_y-1\},
\end{equation}
 The $j$-th RF beamformer $\mathbf{a}_j$ corresponds to a 2D array response that is flattened into a vector. Specifically, the $(n,m)$-th entry of the underlying 2D beam pattern is given by
\begin{equation}
a_j(n,m)
= \exp\!\left(
- j 2\pi
\left( \frac{k_x n}{N_x} + \frac{k_y m}{N_y} \right)
\right),
\end{equation}
with $n = 0,\dots,N_x-1$ and $m = 0,\dots,N_y-1$ denoting the antenna indices along the two array dimensions. The vector $\mathbf{a}_j$ is obtained by stacking $a_j(n,m)$ into a length-$N_t$ column vector using a fixed ordering (i.e., column-wise).

This DFT-based construction yields a set of mutually orthogonal analog beams, satisfying
\begin{equation}
\mathbf{a}_j^{\mathsf{H}} \mathbf{a}_\ell = 0, \qquad j \neq \ell,
\end{equation}
which ensures spatial orthogonality across different codebook entries.

For each user $k$, the analog beam is selected by maximizing the received signal power:
\begin{equation}
\mathbf{a}_{k}^{\star}
= \arg\max_{\mathbf{a}_j \in \mathcal{C}}
\left| \mathbf{h}_k^{\mathsf{H}} \mathbf{a}_j \right|^2,
\end{equation}
where $\mathbf{h}_k \in \mathbb{C}^{N_t \times 1}$ denotes the channel between the BS and user $k$. The analog precoder is then formed by stacking the selected beams:
\begin{equation}
\mathbf{F}_{\mathrm{RF}} =
\begin{bmatrix}
\mathbf{a}_{1}^{\star} &
\mathbf{a}_{2}^{\star} &
\cdots &
\mathbf{a}_{K}^{\star}
\end{bmatrix}.
\end{equation}
This RF precoder provides directional gain through phase-only control, while residual multi-user interference is mitigated in the digital domain in the subsequent step.

\subsubsection{Step 2: Digital Baseband Precoding with Regularized Zero-Forcing}

Given the selected analog precoder $\mathbf{F}_{\mathrm{RF}}$, the effective baseband channel is defined as
\begin{equation}
\mathbf{H}_{\mathrm{eff}} = \mathbf{H}^{\mathsf{H}} \mathbf{F}_{\mathrm{RF}} \in \mathbb{C}^{K \times K},
\end{equation}
where $\mathbf{H} \in \mathbb{C}^{N_t \times K}$ denotes the downlink channel matrix whose columns correspond to the user channels $\{\mathbf{h}_k\}_{k=1}^K$.

To suppress inter-user interference in the digital domain, we employ a regularized zero-forcing (RZF) baseband precoder given by
\begin{equation}
\mathbf{F}_{\mathrm{BB}}
= \beta \,
\mathbf{H}_{\mathrm{eff}}^{\mathsf{H}}
\left(
\mathbf{H}_{\mathrm{eff}} \mathbf{H}_{\mathrm{eff}}^{\mathsf{H}}
+ \alpha \mathbf{I}
\right)^{-1},
\end{equation}
where $\alpha \ge 0$ is the regularization parameter that controls the trade-off between noise amplification and interference suppression, and $\beta$ is a scalar normalization factor chosen to satisfy the transmit power constraint.

The scaling factor $\beta$ is selected such that
\begin{equation}
\left\| \mathbf{F}_{\mathrm{RF}} \mathbf{F}_{\mathrm{BB}} \right\|_F^2 \le P,
\end{equation}
where $P$ denotes the total transmit power budget. In the special case $\alpha = 0$, the proposed digital precoder reduces to the conventional zero-forcing (ZF) solution. This digital precoding stage cancels the residual inter-user interference after analog beamforming, while maintaining robustness to channel conditioning through regularization.

\subsection{Single-User Hybrid Precoding}
\label{app:su_hybrid}

\begin{figure*}[h]
    \centering
\includegraphics[width=1\textwidth]{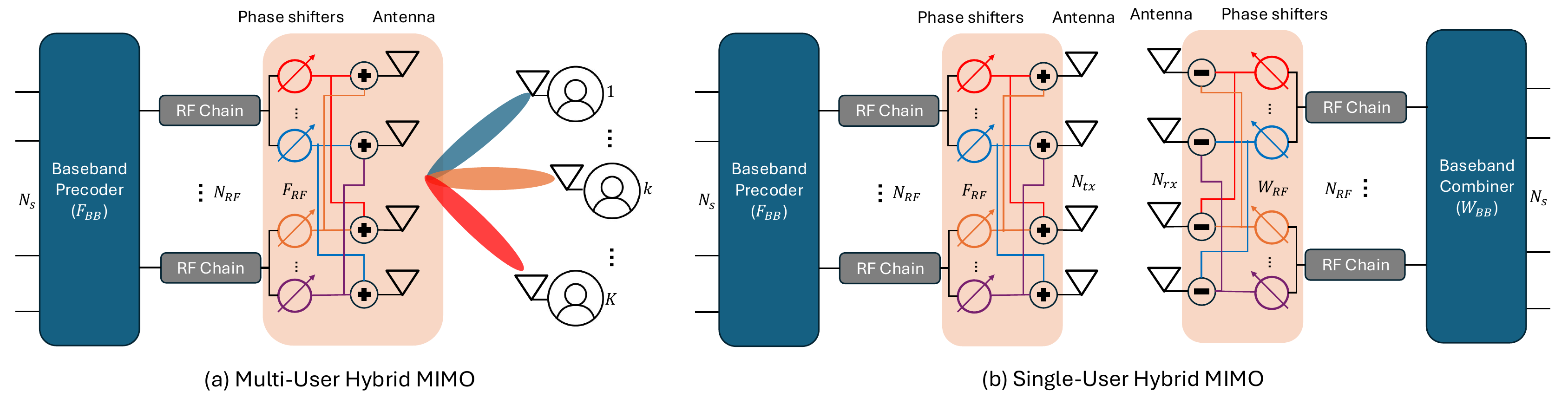} 
\vspace{-5mm}
\caption{\small Hybrid MIMO wireless communication setting in both multi-user and single-user scenarios.}

    \label{fig:MIMO_Setting}
\end{figure*}

We consider a single-user MIMO system in which a transmitter equipped with $N_t$ antennas and $N_{\mathrm{RF}}^{\mathrm{t}}$ RF chains communicates with a receiver equipped with $N_r$ antennas and $N_{\mathrm{RF}}^{\mathrm{r}}$ RF chains, as illustrated in Figure~\ref{fig:MIMO_Setting} (b). The system supports $N_s$ data streams, with $N_s \le \min(N_{\mathrm{RF}}^{\mathrm{t}}, N_{\mathrm{RF}}^{\mathrm{r}})$. The narrowband channel between the transmitter and the receiver is denoted by $\mathbf{H} \in \mathbb{C}^{N_r \times N_t}$.

A hybrid precoding and combining architecture is adopted at both ends. The transmitted signal is
\begin{equation}
\mathbf{x} = \mathbf{F}_{\mathrm{RF}} \mathbf{F}_{\mathrm{BB}} \mathbf{s},
\end{equation}
where $\mathbf{F}_{\mathrm{RF}} \in \mathbb{C}^{N_t \times N_{\mathrm{RF}}^{\mathrm{t}}}$ and $\mathbf{F}_{\mathrm{BB}} \in \mathbb{C}^{N_{\mathrm{RF}}^{\mathrm{t}} \times N_s}$ denote the analog and digital precoders, respectively, and $\mathbf{s} \in \mathbb{C}^{N_s \times 1}$ is the data symbol vector with $\mathbb{E}[\mathbf{s}\mathbf{s}^{\mathsf{H}}] = \mathbf{I}_{N_s}$.  

At the receiver, the signal is first processed by an analog combiner $\mathbf{W}_{\mathrm{RF}} \in \mathbb{C}^{N_r \times N_{\mathrm{RF}}^{\mathrm{r}}}$, and then by a digital combiner $\mathbf{W}_{\mathrm{BB}} \in \mathbb{C}^{N_{\mathrm{RF}}^{\mathrm{r}} \times N_s}$. The resulting processed received signal is
\begin{equation}
\mathbf{y} = \mathbf{W}_{\mathrm{BB}}^{\mathsf{H}} \mathbf{W}_{\mathrm{RF}}^{\mathsf{H}} \mathbf{H} \mathbf{F}_{\mathrm{RF}} \mathbf{F}_{\mathrm{BB}} \mathbf{s} + \mathbf{W}_{\mathrm{BB}}^{\mathsf{H}} \mathbf{W}_{\mathrm{RF}}^{\mathsf{H}} \mathbf{n},
\end{equation}
where $\mathbf{n} \sim \mathcal{CN}(\mathbf{0}, \sigma^2 \mathbf{I}_{N_r})$ represents the additive white Gaussian noise at the receiver antennas.

\subsubsection{Hybrid Precoding Optimization Problem}

The objective is to jointly design the hybrid precoder $\{\mathbf{F}_{\mathrm{RF}}, \mathbf{F}_{\mathrm{BB}}\}$ and hybrid combiner $\{\mathbf{W}_{\mathrm{RF}}, \mathbf{W}_{\mathrm{BB}}\}$ to maximize the achievable spectral efficiency under practical hardware constraints. The resulting optimization problem is formulated as
\begin{equation}
\begin{aligned}
\max_{\mathbf{F}_{\mathrm{RF}},\,\mathbf{F}_{\mathrm{BB}},\,\mathbf{W}_{\mathrm{RF}},\,\mathbf{W}_{\mathrm{BB}}}
\quad &
\log\!\det\!\left(
\mathbf{I}
+
\frac{\mathrm{SNR}}{N_s}
\mathbf{W}_{\mathrm{BB}}^{\mathsf{H}}
\mathbf{W}_{\mathrm{RF}}^{\mathsf{H}}
\mathbf{H}
\mathbf{F}_{\mathrm{RF}}
\mathbf{F}_{\mathrm{BB}}
\mathbf{F}_{\mathrm{BB}}^{\mathsf{H}}
\mathbf{F}_{\mathrm{RF}}^{\mathsf{H}}
\mathbf{H}^{\mathsf{H}}
\mathbf{W}_{\mathrm{RF}}
\mathbf{W}_{\mathrm{BB}}
\right) \\
\text{s.t.} \quad &
\left[ \mathbf{F}_{\mathrm{RF}} \right]_{(:,n_t)} \in \mathcal{F}, \quad \forall n_t, \\
& \left[ \mathbf{W}_{\mathrm{RF}} \right]_{(:,n_r)} \in \mathcal{W}, \quad \forall n_r, \\
& \| \mathbf{F}_{\mathrm{RF}} \mathbf{F}_{\mathrm{BB}} \|_F^2 \le P,
\end{aligned}
\end{equation}
where $\mathcal{F}$ and $\mathcal{W}$ denote the transmit and receive analog beam codebooks, and $P$ is the total transmit power budget.

This joint optimization problem is highly non-convex due to the constant-modulus constraints imposed on the analog beamformers and combiners, as well as the coupled nature of the analog and digital variables. To obtain a tractable solution, we adopt a two-step hybrid design approach assuming orthogonal vectors in TX/RX codebooks. In the first step, the analog beamformer and combiner are selected via a greedy exhaustive search over the beam codebooks, assuming no digital processing. In the second step, the digital baseband precoder and combiner are designed based on the resulting effective channel. It should be noted that, while this approach attains an upper-bound performance under full channel state information, it is impractical for real-time deployment due to the prohibitive complexity of the exhaustive search, which scales with the number of antennas (i.e., the codebook size). We use this baseline as the upper bound performance that can be achieved by the AI-driven solutions. 

\subsubsection{Step 1: Analog Beamformer and Combiner Selection}

In the first step, the analog precoder and combiner are jointly selected from predefined beam codebooks using a greedy exhaustive search. The goal is to maximize the achievable spectral efficiency assuming only analog processing. Specifically, the optimal analog beamformer and combiner are obtained as
\begin{equation}
\{\mathbf{F}_{\mathrm{RF}}, \mathbf{W}_{\mathrm{RF}}\}
=
\arg\max_{\mathbf{F}_{\mathrm{RF}},\,\mathbf{W}_{\mathrm{RF}}}
\log\!\det\!\left(
\mathbf{I}
+
\frac{\mathrm{SNR}}{N_s}
\mathbf{W}_{\mathrm{RF}}^{\mathsf{H}}
\mathbf{H}
\mathbf{F}_{\mathrm{RF}}
\left( \mathbf{F}_{\mathrm{RF}}^{\mathsf{H}} \mathbf{F}_{\mathrm{RF}} \right)^{-1}
\mathbf{F}_{\mathrm{RF}}^{\mathsf{H}}
\mathbf{H}^{\mathsf{H}}
\mathbf{W}_{\mathrm{RF}}
\left( \mathbf{W}_{\mathrm{RF}}^{\mathsf{H}} \mathbf{W}_{\mathrm{RF}} \right)^{-1}
\right),
\end{equation}
subject to the constraints
\begin{equation}
\left[ \mathbf{F}_{\mathrm{RF}} \right]_{(:,n_t)} \in \mathcal{F},
\quad \forall n_t,
\qquad
\left[ \mathbf{W}_{\mathrm{RF}} \right]_{(:,n_r)} \in \mathcal{W},
\quad \forall n_r,
\end{equation}
where $\mathcal{F}$ and $\mathcal{W}$ denote the transmit and receive analog beam codebooks, respectively.

Both codebooks are constructed using DFT-based steering vectors for uniform planar arrays, as described in MU-MIMO framework~\ref{app:mu_hybrid}. This construction yields constant-modulus, mutually orthogonal beams spanning discrete angular directions. \textbf{We design a greedy exhaustive search that evaluates all feasible combinations of beamformer and combiner indices and selects the pair that maximizes the above objective and provides the optimal labels.}

\subsubsection{Step 2: Digital Baseband Precoding and Combining}

Given the selected analog beamformer and combiner, the effective low-dimensional channel is
\begin{equation}
\mathbf{H}_{\mathrm{eff}} = \mathbf{W}_{\mathrm{RF}}^{\mathsf{H}} \mathbf{H} \mathbf{F}_{\mathrm{RF}}.
\end{equation}
The effective channel is then decomposed using singular value decomposition (SVD) as
\begin{equation}
\mathbf{H}_{\mathrm{eff}} = \mathbf{U}_s \boldsymbol{\Sigma}_s \mathbf{V}_s^{\mathsf{H}},
\end{equation}
where $\mathbf{U}_s$ and $\mathbf{V}_s$ contain the left and right singular vectors corresponding to the $N_s$ dominant singular values in $\boldsymbol{\Sigma}_s$.

The digital precoder and combiner are designed to align with the dominant singular modes of the effective channel:
\begin{equation}
\mathbf{F}_{\mathrm{BB}} = \mathbf{F}_{\mathrm{RF}}^{-1} \mathbf{V}_s,
\qquad
\mathbf{W}_{\mathrm{BB}} = \mathbf{W}_{\mathrm{RF}}^{-1} \mathbf{U}_s.
\end{equation}
Finally, $\mathbf{F}_{\mathrm{BB}}$ is appropriately normalized to satisfy the transmit power constraint. This two-step hybrid design leverages directional gain through analog beamforming and achieves near-optimal spatial multiplexing performance via digital baseband processing.

\section{Pretraining Recounstruction Objectives Visualization}
\label{app:pretraining_objectives}
In this section, we visualize the pretraining objectives of the foundational model by examining how it attempts to reconstruct cross-modality information from wireless channel observations. Specifically, we illustrate the model’s predictions on examples of previously unseen wireless channel conditions. The considered reconstruction tasks include: (1) user position estimation from partial CSI, (2) spatial spectrum reconstruction from partial CSI, and (3) occupancy grid reconstruction from full CSI combined with location information.

\begin{figure*}[h]
    \centering
\includegraphics[width=0.95\textwidth]{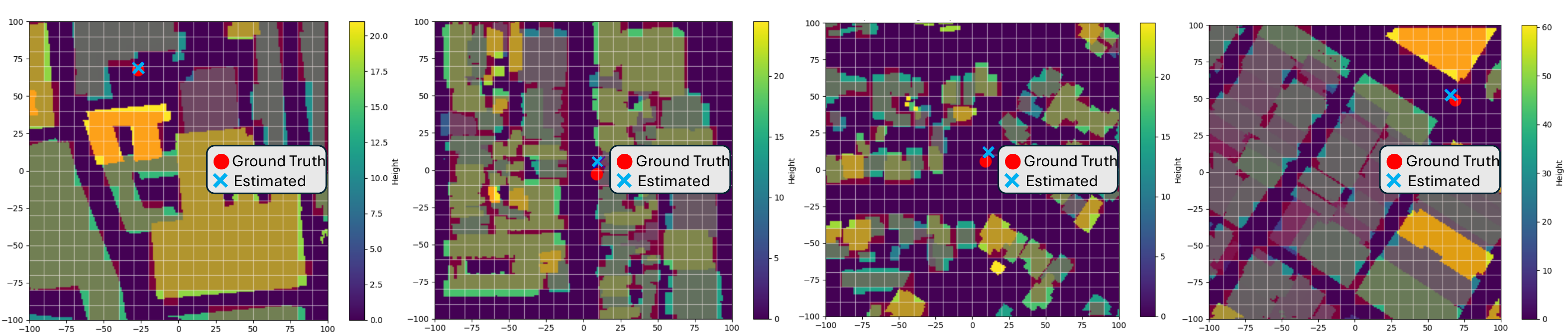} 
\vspace{-3mm}
\caption{User location reconstruction from partial channel state information and scene information.}
\vspace{-3mm}
    \label{fig:pre_loc}
\end{figure*}

\begin{figure*}[h]
    \centering
\includegraphics[width=1\textwidth]{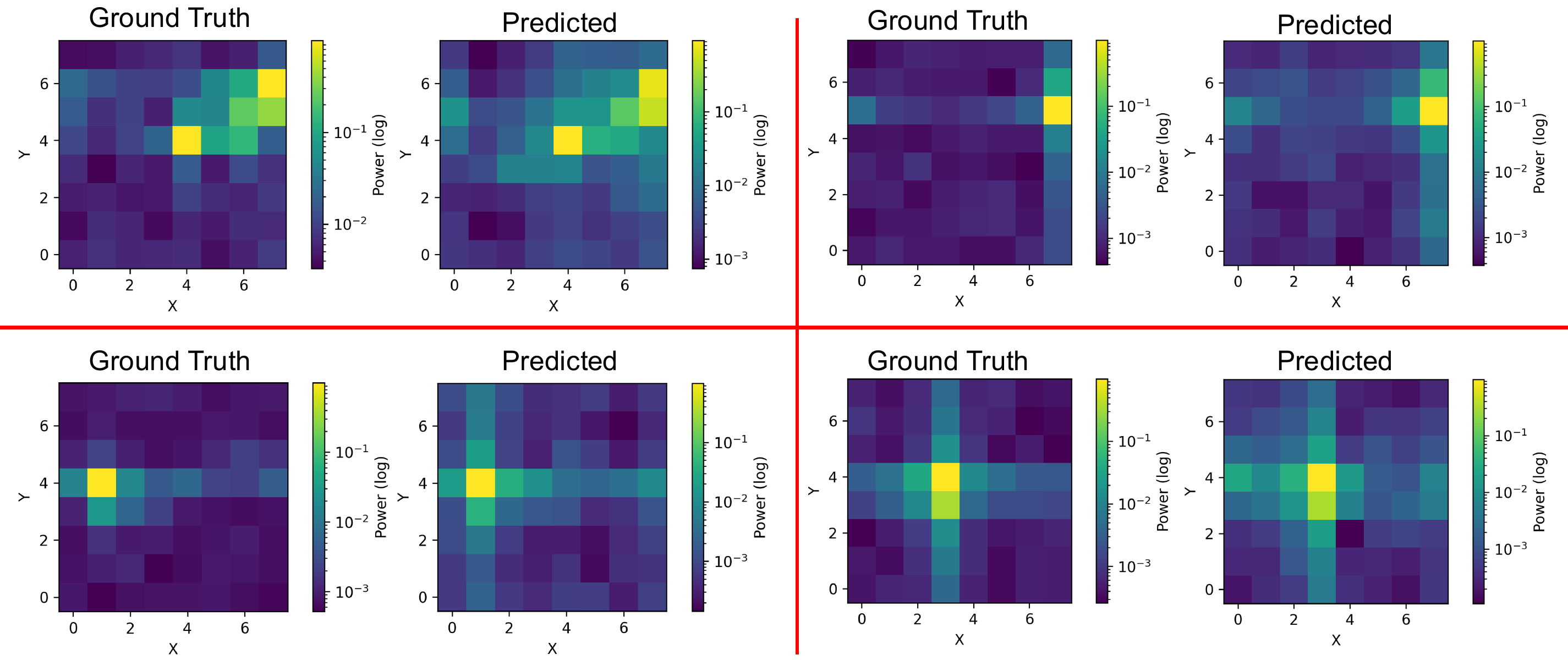} 
\vspace{-5mm}
\caption{Spatial spectra reconstruction from partial channel state information and scene information.}
    \label{fig:pre_spatial}
\end{figure*}

Figure~\ref{fig:pre_loc} presents four deployment areas, showing the ground-truth user locations together with the corresponding positions reconstructed from partial CSI and 3D static scene information by the model. These examples provide a qualitative visualization of the model's attempt to relate structural patterns observed in partial CSI to user location, by exploiting correlations induced by the underlying propagation environment. Figure~\ref{fig:pre_spatial} compares the ground-truth and predicted spatial spectra for four unseen samples reconstructed from partial CSI, illustrating the model's attempt to recover dominant spatial features of the propagation channel that guides the attention pattern to encode physically relevant information.

\begin{figure*}[h]
    \centering
\includegraphics[width=1\textwidth]{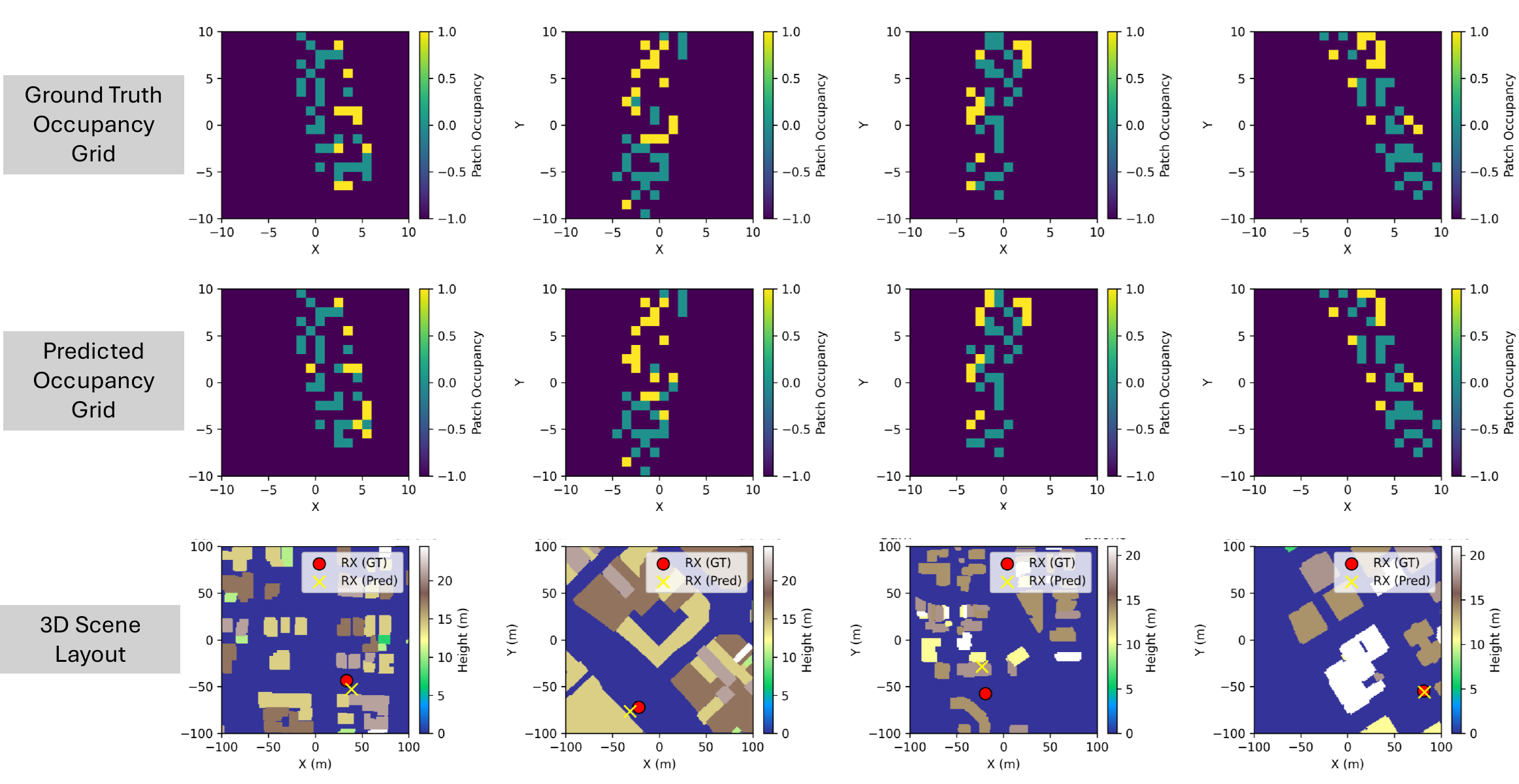}
\vspace{-6mm}
\caption{Occupancy grid reconstruction from wireless channel state information and user location.}
    \label{fig:pre_scene}
\end{figure*}

Finally, Figure~\ref{fig:pre_scene} illustrates four deployment scenarios along with the corresponding ground-truth occupancy grids and the predicted grids reconstructed from the CSI of a particular receiver, which is marked by a red circle in the scene layout. The original deployment area of size $200,\mathrm{m} \times 200,\mathrm{m}$ is discretized into $10,\mathrm{m} \times 10,\mathrm{m}$ patches, resulting in an occupancy grid of size $20 \times 20$. The masking strategy employed during pretraining is also visualized, where only a subset of patches within a predefined distance threshold from the line connecting the base station and the receiver is masked. Unmasked patches are indicated by a value of $-1$, while masked patches take binary values of 0 and 1, representing the estimated or ground-truth occupancy of the corresponding regions. These visualizations are intended to qualitatively illustrate how the model attempts to encode bidirectional cross-modality associations through its pretraining objectives, by mapping observable structures in CSI, reflecting line-of-sight, non-line-of-sight, and diffraction effects, to latent representations of spatial and environmental properties.

\clearpage
\section{Details on Downstream Task Implementation} 
\label{app:downstream_task_imp}
\subsection{Localization} 
\label{app:localization_setting}
\begin{figure*}[h]
    \centering
\includegraphics[width=1\textwidth]{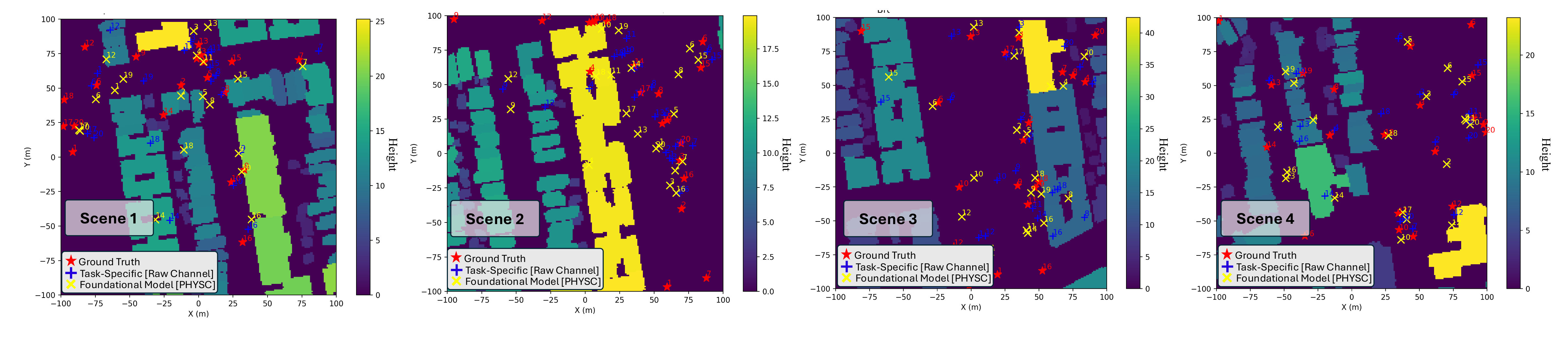}
\caption{Visualization of the 3D scene layouts in the deployment areas of four test scenes for the downstream localization task, showing 20 user location samples along with the corresponding predictions from the foundational model and the task-specific model.}
    \label{fig:localization_layout}
\end{figure*}

Figure~\ref{fig:localization_layout} illustrates the experimental setup for the downstream localization task. It visualizes the 3D layouts of the deployment areas for four distinct test scenes, highlighting the structural characteristics of each environment, including building properties in the outdoor deployment area and the corresponding height at each spatial location. The figure further illustrates the predicted user locations obtained from the foundational model and the task-specific localization model with the ground-truth positions for 20 randomly selected user locations, under a zero-shot cross-scene generalization setting.

\subsubsection{Training settings.}
We consider two localization scenarios: \emph{scene-specific localization} and \emph{cross-scene localization}. In both cases, we evaluate two models: (i) a task-specific baseline trained directly on CSI, and (ii) the proposed foundational model equipped with a lightweight localization projection head. The task-specific baseline  operates on the complex CSI represented by two channels (real and imaginary) and consists of three convolutional blocks with channel dimensions $\{16,32,64\}$, each followed by batch normalization and ReLU activations. Spatial resolution is progressively reduced via max pooling, and global features are extracted using adaptive average pooling. A shallow regressor (64$\rightarrow$32$\rightarrow$2) predicts the 2D receiver coordinates, followed by a $\tanh$ activation to constrain outputs to the environment range. For the foundational model, localization is performed by attaching a lightweight MLP projection head to the encoded $[\mathrm{PHYSC}]$ token, implemented as a three-layer perceptron (128$\rightarrow$64$\rightarrow$64$\rightarrow$2) with ReLU activations and a final $\tanh$ nonlinearity, after which predictions are scaled to the physical environment dimensions.

For scene-specific localization, both projection models are trained from scratch using a learning rate of $10^{-4}$. For cross-scene localization, models are trained on source scenes and then fine-tuned on the target scene using a learning rate of $10^{-5}$ to encourage transfer while preserving pretrained representations. This unified setting enables a direct comparison between task-specific learning and foundation-model-based adaptation under both in-scene and cross-scene generalization.

\subsection{MU-MIMO Precoding}
\label{app:MU_MIMO_Imp}
\subsubsection{Achieved Sum-Rate} 
\label{app:sum-rate_ecdf}
\begin{figure*}[h]
    \centering
\includegraphics[width=1\textwidth]{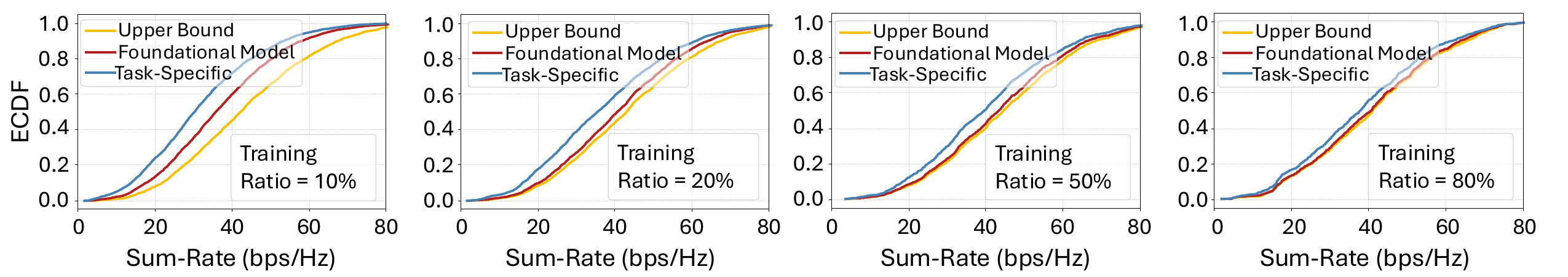}
\vspace{-3mm}
\caption{Empirical cumulative distribution function (ECDF) of the achieved sum-rate for all user clusters in the test set, comparing the foundational model, the task-specific model, and the upper bound across different training ratios.}
    \label{fig:ecdf_sum_rate}
\end{figure*}

Figure~\ref{fig:ecdf_sum_rate} illustrates the empirical cumulative distribution function (ECDF) of the achieved sum-rate for all user clusters in the test set, comparing the foundational model leveraging the $[\mathrm{PHYSC}]$ token, the task-specific model trained with raw channel information, and the upper-bound performance, across different training ratios representing the available labeled data. The results clearly show that the foundational model consistently outperforms the task-specific model, particularly in data-limited scenarios, and substantially improves the achievable data rate, approaching the upper-bound performance with limited training data. This demonstrates the efficiency of the foundational model in learning meaningful and universal representations that can effectively adapt to wireless communication tasks, achieving near-optimal data-rate performance with nearly four times fewer labeled samples.

\subsubsection{Classification Head Training} 
\label{app:cls_head}
\begin{figure*}[h]
    \centering
\includegraphics[width=1\textwidth]{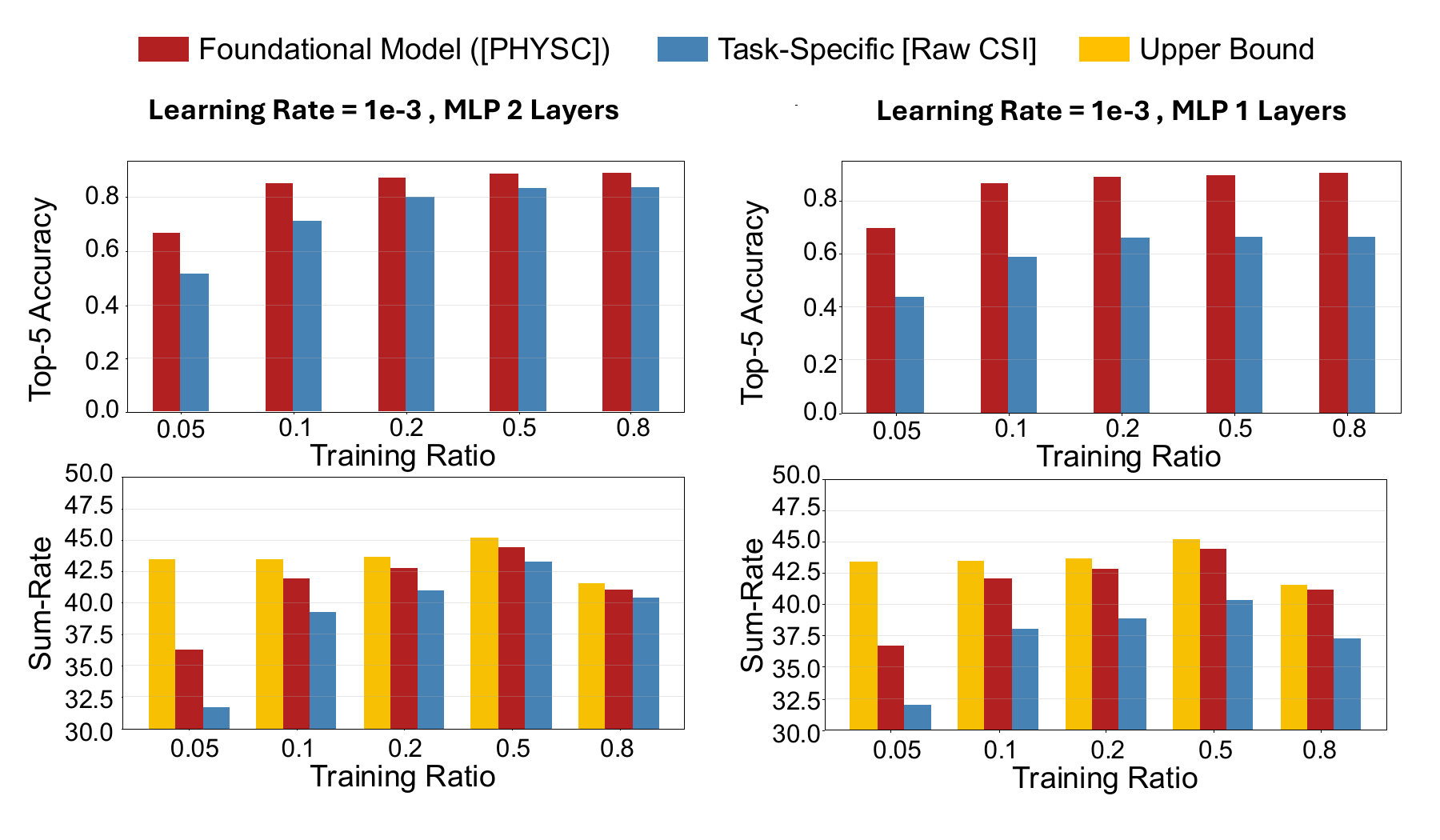}
\caption{Ablation study on the impact of MLP head configuration for beamformer index prediction, showing the average top-5 beam detection accuracy and achieved sum-rate for the foundational model and the task-specific model across different training ratios, hidden-layer configurations, with learning rate = 1e-3.}
    \label{fig:mimo_ablation1}
\end{figure*}

\begin{figure*}[t]
    \centering
\includegraphics[width=1\textwidth]{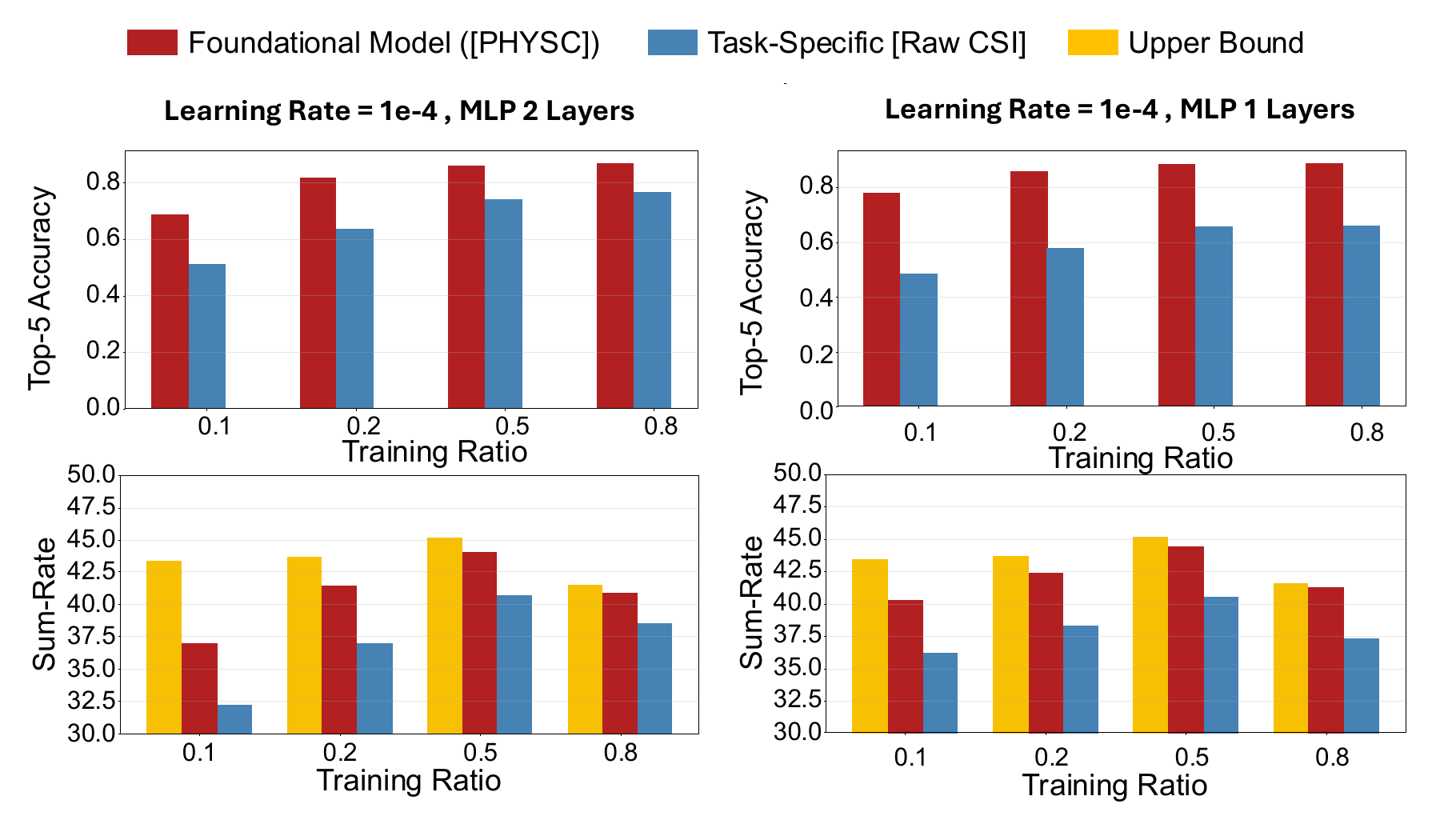}
\caption{Ablation study on the impact of MLP head configuration for beamformer index prediction, showing the average top-5 beam detection accuracy and achieved sum-rate for the foundational model and the task-specific model across different training ratios, hidden-layer configurations, with learning rate = 1e-4.}
    \label{fig:mimo_ablation2}
\end{figure*}

As described earlier, to identify the optimal beamformer index from the codebook, we leverage the encoded $[\mathrm{PHYSC}]$ token produced by the foundational model and feed it into an MLP head, which is trained using a weighted cross-entropy loss to predict the dominant propagation path. For a fair comparison, we employ a similar MLP architecture for the task-specific model, which operates directly on raw channel information and is trained with the same objective. Since the input dimensionality to the classification head is identical (128) for both models, the two frameworks have similar number of learnable parameters. We evaluate the impact of classification head complexity by varying the MLP configuration, using different learning rates and number of hidden layers of size 32. Figure~\ref{fig:mimo_ablation1} and ~\ref{fig:mimo_ablation2} reports the average top-5 beam detection accuracy and the achieved sum-rate across different training ratios for the considered MLP configurations. The results consistently demonstrate the superior performance of the foundational model across all settings, with particularly pronounced gains in data-limited regimes.

%%%%%%%%%%%%%%%%%%%%%%%%%%%%%%%%%%%%%%%%%%%%%%%%%%%%%%%%%%%%%%%%%%%%%%%%%%%%%%%
%%%%%%%%%%%%%%%%%%%%%%%%%%%%%%%%%%%%%%%%%%%%%%%%%%%%%%%%%%%%%%%%%%%%%%%%%%%%%%%

\end{document}